\documentclass[%
preprint,
amsmath,amssymb,
aps]{revtex4-1}
\usepackage{txfonts}
\usepackage{amsmath,amssymb}
\usepackage{comment}
\usepackage[dvipdfmx]{graphicx}
\usepackage{color}
\usepackage{dcolumn}
\usepackage{bm}
\usepackage{url}
\usepackage{braket}
\usepackage{setspace}
\usepackage{ascmac}

\usepackage{hyperref} 
\hypersetup{ 
    colorlinks=true,       
    linkcolor=cyan,          
    citecolor=magenta,        
    filecolor=magenta,      
    urlcolor=blue,           
    runcolor=cyan  
} 
 
\begin{document}


\title{Reverse annealing for the fully connected $p$-spin model}

\author{Masaki Ohkuwa}
\email{ookuwa.m@stat.phys.titech.ac.jp}
\author{Hidetoshi Nishimori}%
\affiliation{%
Department of Physics, Tokyo Institute of Technology, Oh-okayama, Meguro-ku, Tokyo 152-5551, Japan 
}%
\author{Daniel A. Lidar}
\affiliation{Departments of Electrical Engineering, Chemistry, and Physics \& Astronomy, University of Southern California, Los Angeles, California 90089, USA}
 \affiliation{Center for Quantum Information Science \& Technology, University of Southern California, Los Angeles, California 90089, USA}

\date{\today}%

\begin{abstract}

Reverse annealing is a variant of quantum annealing that starts from a given classical configuration of spins (qubits). In contrast to the conventional formulation, where one starts from a uniform superposition of all possible states (classical configurations), quantum fluctuations are first increased and only then decreased. One then reads out the state as a proposed solution to the given combinatorial optimization problem. We formulate a mean-field theory of reverse annealing using the fully-connected ferromagnetic $p$-spin model, with and without random longitudinal fields, and analyze it in order to understand how and when reverse annealing is effective at solving this problem. We find that the difficulty arising from the existence of a first-order quantum phase transition, which leads to an exponentially long computation time in conventional quantum annealing, is circumvented in the context of this particular problem by reverse annealing if the proximity of the initial state to the (known) solution exceeds a threshold.  Even when a first-order transition is unavoidable, the difficulty is mitigated due to a smaller jump in the order parameter at a first-order transition, which implies a larger rate of quantum tunneling. This is the first analytical study of reverse annealing and paves the way toward a systematic understanding of this relatively unexplored protocol in a broader context.

\end{abstract}

\pacs{Valid PACS appear here}
\maketitle

\section{\label{sec:level1}Introduction\protect}

Quantum annealing (QA) is a metaheuristic designed to solve combinatorial optimization problems by exploiting quantum fluctuations \cite{Kadowaki1998,Brooke1999,Santoro2002,Santoro2006,Das2008,Morita2008,Tanaka_book2017}, and is closely related to adiabatic quantum computing \cite{Farhi2000,Farhi2001,AlbashLidar2018}.  Combinatorial optimization is a class of problems in which the goal is to find the global minimum of a cost function of many discrete variables. The cost function can in many cases of interest be expressed  as the Hamiltonian of an Ising model with long-range two-body interactions \cite{Lucas2014}. We can therefore make use of the toolbox of statistical mechanics to study combinatorial optimization problems.

The process of conventional QA starts from the uniform superposition of all possible classical states, which is the ground state of a uniform transverse field. One then gradually decreases the amplitude of the transverse field toward zero, and the final state is expected to be the solution to the given combinatorial optimization problem. In contrast, the interesting method of reverse annealing, proposed and first tested by Perdomo-Ortiz {\em et al.} \cite{Perdomo2011} under the name of "sombrero adiabatic quantum computation", starts from a candidate state expected to be closer to the final solution than a random guess.  One then follows the two-stage process of an increase and then a decrease of the amplitude of transverse field.  As shown numerically in Ref. \cite{Perdomo2011}, this approach can lead to better results if the initial state is appropriately chosen. This feature has been implemented in the latest model of the D-Wave device, and used successfully in the context of a quantum simulation experiment \cite{King:2018ab}. Reverse annealing can be viewed as a member of a larger family of performance enhancement methods for quantum annealing via path modification~\cite{AlbashLidar2018}.

The goal of the present paper is to establish an analytical framework to study the performance of reverse annealing through a mean-field theory.  We formulate the problem in terms of the infinite-range many-body-interacting $p$-spin model and study its thermodynamic properties. The result makes it possible to predict, in a static setting, whether or not the difficulties in conventional QA can be removed, or at least mitigated, by reverse annealing. 

In the next Section, we formulate and solve the mean-field theory of reverse annealing for the $p$-body interacting system with and without longitudinal random fields.  The method is further applied to the case of non-stoquastic Hamiltonians in Sec.~\ref{sec:non-stoquastic}.  We conclude in Section~\ref{sec:conclusions}, and provide some additional technical details in the Appendixes.


\section{Reverse annealing for the $p$-spin model}

We first formulate the problem and proceed to the description of the results of the statistical-mechanical analysis.

\subsection{Formulation}

Let us describe reverse annealing by the following Hamiltonian,
\begin{align}
\label{eq:hamiltonian1}
\hat{H}(s,\lambda)=s\hat{H}_0+(1-s)(1-\lambda)\hat{H}_{{\rm init}}+(1-s)\lambda\hat{V}_{{\rm TF}}\quad (0\le s, \lambda \le 1),
\end{align}
where $\hat{H}_0$ is the target Hamiltonian, $\hat{H}_{{\rm init}}$ determines the initial state, and $\hat{V}_{{\rm TF}}$ denotes the transverse field. We choose the following forms of these terms,
\begin{align}
\hat{H}_0=-N\left(\frac{1}{N}\sum_{i=1}^N\hat{\sigma}_i^z\right)^p-\sum_{i=1}^Nh_i\hat{\sigma}_i^z,\quad
\hat{H}_{{\rm init}}=-\sum_{i=1}^N\epsilon_i\hat{\sigma}_i^z,\quad
\hat{V}_{{\rm TF}}=-\sum_{i=1}^N\hat{\sigma}_i^x,
\end{align}
where $p$ is a positive integer, the $\hat{\sigma}_i$ are the usual Pauli operators at site $i$, and $N$ is the number of spins (qubits).
Conventional quantum annealing is reproduced with $\lambda=1$, in which case $\hat{H}_{\rm init}$ drops out of the Hamiltonian. The initial values of the parameters are $s=\lambda=0$, upon which only the second term on the right-hand side of Eq.~(\ref{eq:hamiltonian1}) remains and the ground state is set to the initial state, $\hat{\sigma}_i^z=\epsilon_i~(\forall i)$, where $\epsilon_i~(=\pm 1)$ is a candidate solution expected to be close to the correct ground state.
We next let the system evolve adiabatically toward the goal of $s=\lambda =1$, where only the target Hamiltonian $\hat{H}_0$ survives. We consider local field variables $h_i$ that are either zero, random bimodal $h_i=\pm h_0$, or Gaussian-distributed.

When $p\ge 3$ and $\lambda=1$, this model is known to undergo a first-order transition as a function of $s$ \cite{Jorg2010}. We choose $p$ to be odd to avoid the trivial double degeneracy for even $p$, except for the interesting case of $p=2$, which corresponds to two-body interactions.

\subsection{Results}

We follow the standard procedure for evaluating the partition function by the Suzuki-Trotter decomposition and the static approximation, almost in the same way as described in Ref.~\cite{Seki2012}. See also \footnote{Manaka Okuyama (private communication) recently proved that the static approximation gives the exact solution in the present problem. We nevertheless use the term `static approximation' following the convention.}. The resulting free energy as a function of magnetization $m~\big(=\langle \sum_{i=1}^N \hat{\sigma}_i^z\rangle /N\big)$ is
\begin{align}
f=s(p-1)m^p-T\left[\ln2\cosh\beta\sqrt{\left(spm^{p-1}+sh_i+(1-s)(1-\lambda)\epsilon_i\right)^2+(1-s)^2\lambda^2}\right]_i,
\label{eq:free_energy_Tfinite}
\end{align}
where $\beta$ is the inverse temperature $\beta=1/T$ and the brackets $[\cdots ]_i$ stand for the average over sites
\begin{equation}
    \Big[(\cdots)\Big]_i =\frac{1}{N}\sum_{i=1}^N (\cdots).
\end{equation}
In the low-temperature limit $T\to 0$, the free energy and its minimization condition, the self-consistent equation, are
\begin{align}
f=s(p-1)m^p-\left[\sqrt{(spm^{p-1}+sh_i+(1-s)(1-\lambda)\epsilon_j)^2+(1-s)^2\lambda^2}\right]_i
\label{eq:free_energy_T=0}
\end{align}
and
\begin{align}
\label{eq:self_consistent_eq}
m=\left[\frac{spm^{p-1}+sh_i+(1-s)(1-\lambda)\epsilon_i}{\sqrt{(spm^{p-1}+sh_i+(1-s)(1-\lambda)\epsilon_i)^2+(1-s)^2\lambda^2}}\right]_i,
\end{align}
respectively.
\subsubsection{No random field}\label{subsubsection:NoRandomField1}

Let us first study the simplest case of no random field ($h_i=0,~\forall i)$. The free energy and the self-consistent equation reduce to
\begin{align}
\label{free_energy_zero}
f=s(p-1)m^p&-c\sqrt{(spm^{p-1}+(1-s)(1-\lambda))^2+(1-s)^2\lambda^2}\nonumber \\
&-(1-c)\sqrt{(spm^{p-1}-(1-s)(1-\lambda))^2+(1-s)^2\lambda^2},
\end{align}
and
\begin{align}
\label{eq:self_consist_zero}
m=&c\frac{spm^{p-1}+(1-s)(1-\lambda)}{\sqrt{(spm^{p-1}+(1-s)(1-\lambda))^2+(1-s)^2\lambda^2}}\nonumber \\
&+(1-c)\frac{spm^{p-1}-(1-s)(1-\lambda)}{\sqrt{(spm^{p-1}-(1-s)(1-\lambda))^2+(1-s)^2\lambda^2}},
\end{align}
where $c\;(0\le c\le 1)$ is the fraction of the up-spin configuration $(\epsilon_i=1)$ in the initial state, i.e.,
\begin{equation}
c= \frac{1}{N}\sum_{i=1}^N \delta_{\epsilon_i,1} 
\label{eq:c}
\end{equation}
($Nc$ sites have $\epsilon_i=1$). Since the correct ground state of the target Hamiltonian $\hat{H}_0$ has all spins up, a larger value of $c$ means a closer initial state to the correct ground state.

It is known that this model with $\lambda=1$ (conventional QA) has a first-order transition for $p\ge 3$ \cite{Jorg2010}. On the other hand, when $\lambda$ is fixed to 0, a simple analysis of Eq.~(\ref{eq:self_consist_zero}) shows that the magnetization jumps from $2c-1$ to 1 at a critical value $s=s_{\rm c}$, which is determined by the condition $f(2c-1)=f(1)$ as
\begin{align}
\label{eq:zero_transition_point}
s_{\rm c}=\frac{2(1-c)}{1-(2c-1)^p+2(1-c)}\;\;\;\;\;(\lambda=0).
\end{align}

\begin{figure*}[thb]
\centering
    \begin{tabular}{c}
      \begin{minipage}{0.33\linewidth}
			\centering
          \includegraphics[width=5.5cm,clip]{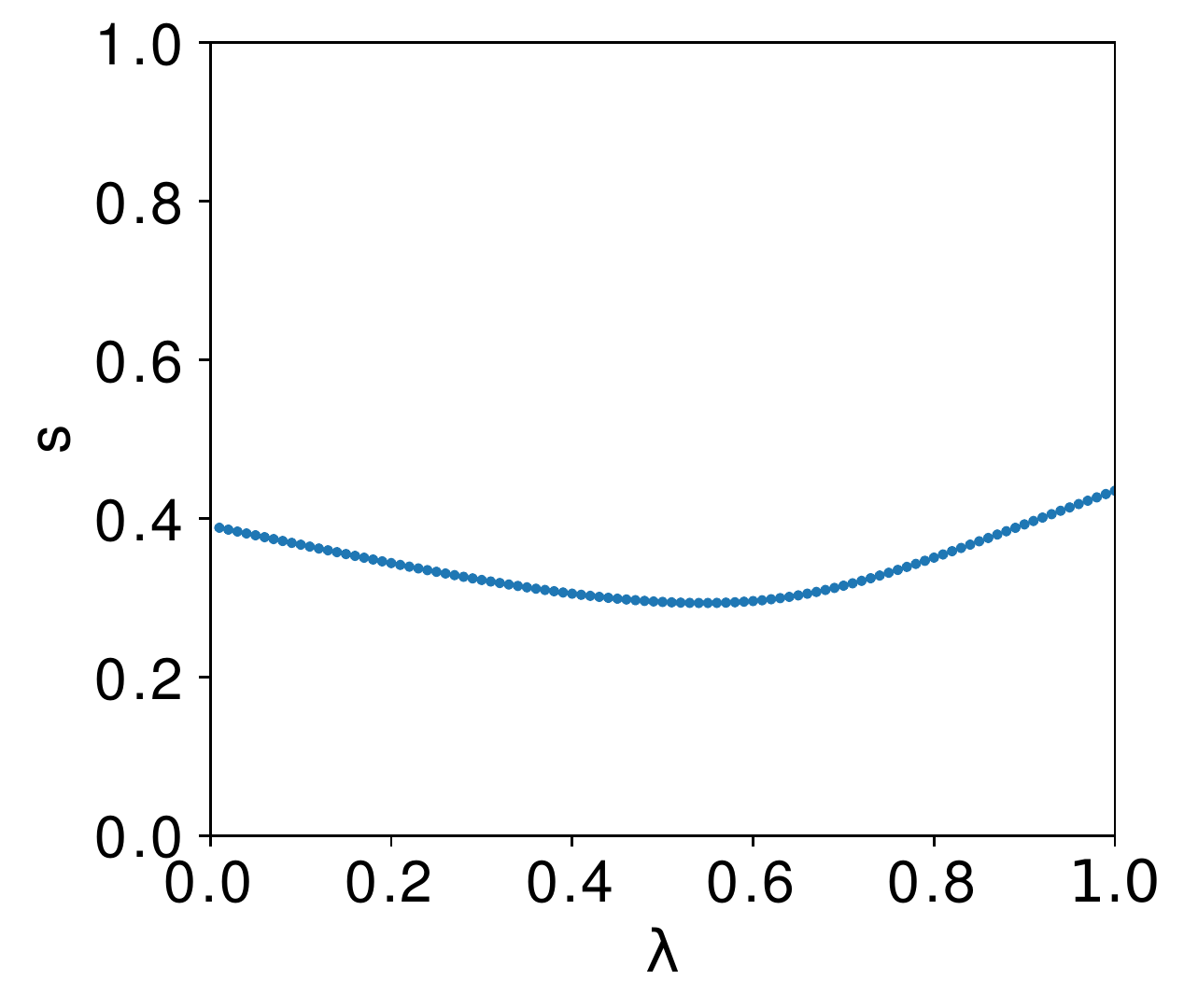}
          \hspace{1.6cm} (a)$\;c=0.7$
      \end{minipage}
      \begin{minipage}{0.33\linewidth}
        \begin{center}
          \includegraphics[width=5.5cm,clip]{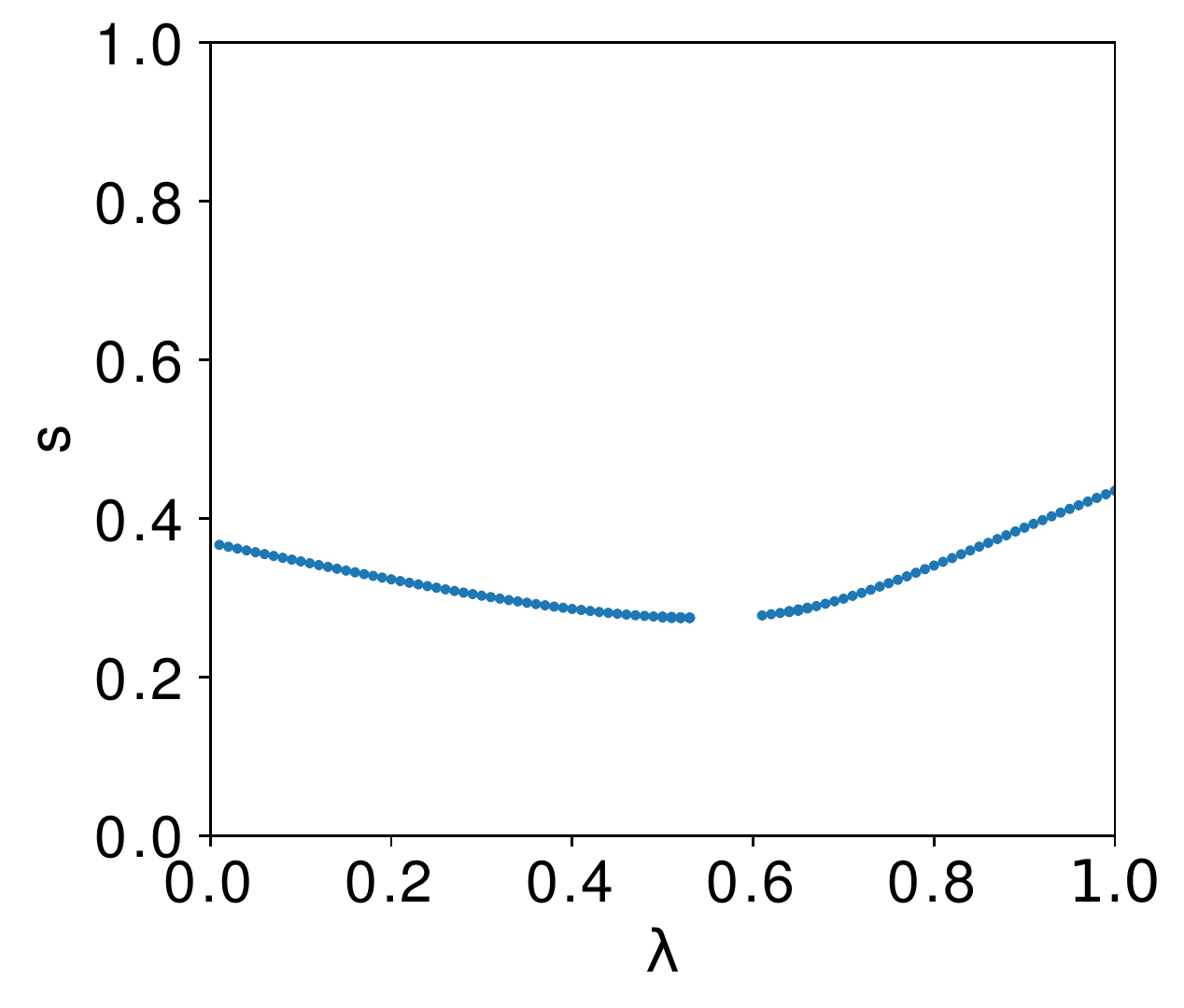}
          \hspace{1.6cm} (b)$\;c=0.74$
        \end{center}
      \end{minipage}
	\begin{minipage}{0.33\linewidth}
			\centering
          \includegraphics[width=5.5cm,clip]{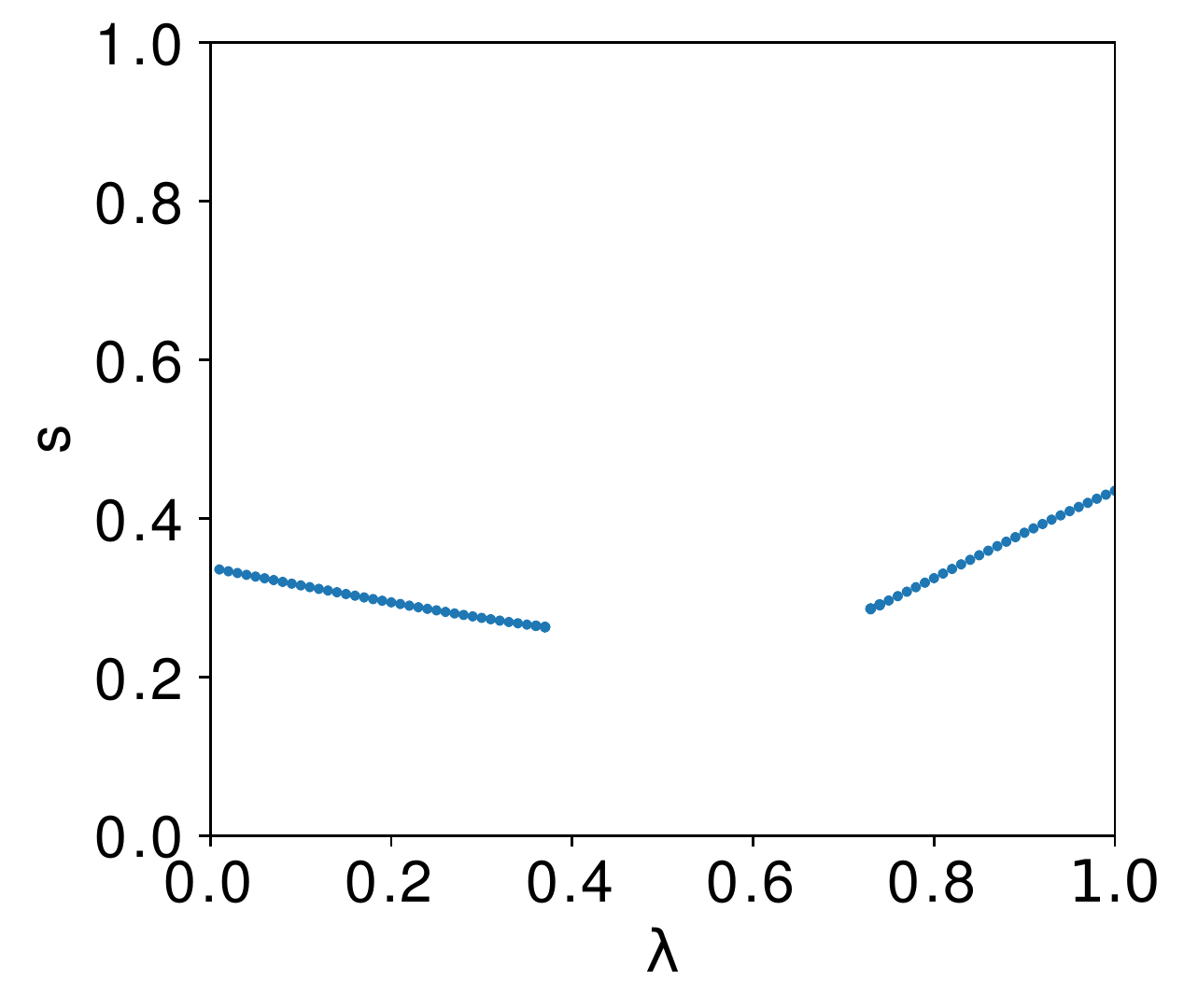}
          \hspace{1.6cm} (c)$\;c=0.8$
      \end{minipage}
    \end{tabular}
    \caption{(Color online) Phase diagrams on the $\lambda$-$s$ plane for $p=3$ and three values of $c$.  The curves represent lines of first-order transitions.}
\label{fig:PDNoRandomField}
\end{figure*}

By numerical evaluation of the free energy and the self-consistent equation, Eqs. (\ref{free_energy_zero}) and (\ref{eq:self_consist_zero}), we obtain the phase diagram in Fig. \ref{fig:PDNoRandomField}, where we chose $p=3$ and (a) $c=0.7$, (b) 0.74, and (c) 0.8. Each curve represents a line of first-order transitions, which is broken at intermediate values of $\lambda$ for $c=0.74$ and 0.8. It is therefore possible to reach the final state $s=\lambda=1$ from the initial state at $s=\lambda =0$ by following a path that avoids first-order transitions. This is to be contrasted with conventional quantum annealing $(\lambda =1)$, where there is no way to avoid a first-order transition.

\begin{figure*}[thb]
\centering
\includegraphics[width=6.5cm,clip]{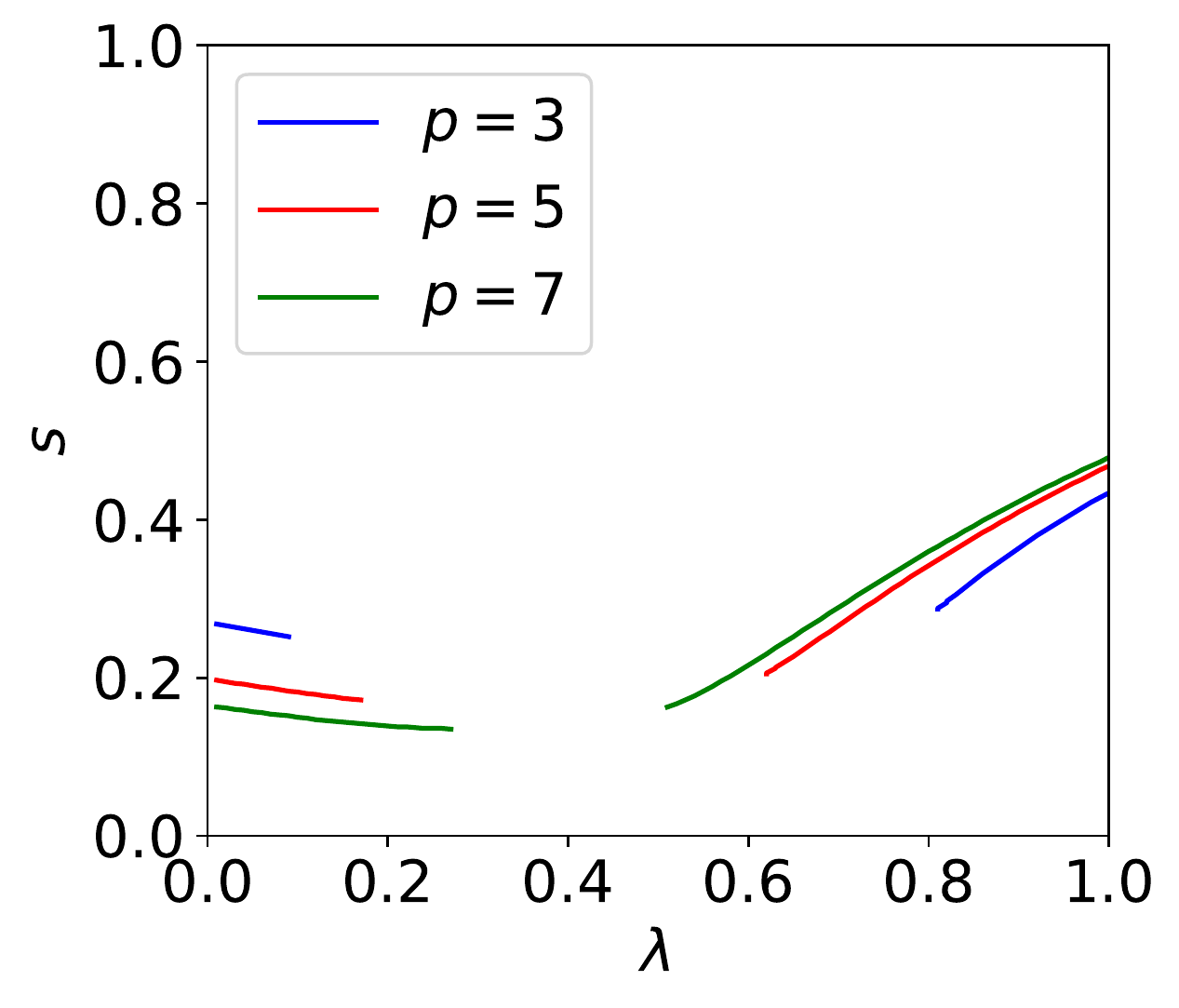}
\caption{(Color online) Phase diagram for $p=3$, 5, and 7. The parameter $c$ is fixed to 0.95.}
\label{fig:357_no_random}
\end{figure*}

Figure \ref{fig:357_no_random} is the phase diagram for $p=3$, 5, and 7 with $c=0.95$. This figure shows that there is a path to avoid first-order transitions for these values of $p$ if $c$ is sufficiently large, though the break in the line of first-order transitions becomes narrower for larger $p$.
Table~\ref{table:p_c_value_noRandom} lists critical values of $c$, beyond which the first-order transition line has a break.
\begin{table}[htb]
	\caption{Critical values of $c$, beyond which a break in the line of first-order transitions shows up in the case without random field.}
  \begin{tabular}{|l||c|c|c|c|} \hline
    $p$ & 3 & 5 & 7 & 11 \\ \hline
    $c$ & 0.74 & 0.89 & 0.94 & 0.97 \\ \hline
  \end{tabular}
	\label{table:p_c_value_noRandom}
\end{table}

\begin{figure*}[thb]
\centering
    \begin{tabular}{c}
      \begin{minipage}{0.5\linewidth}
			\centering
          \includegraphics[width=8cm,clip]{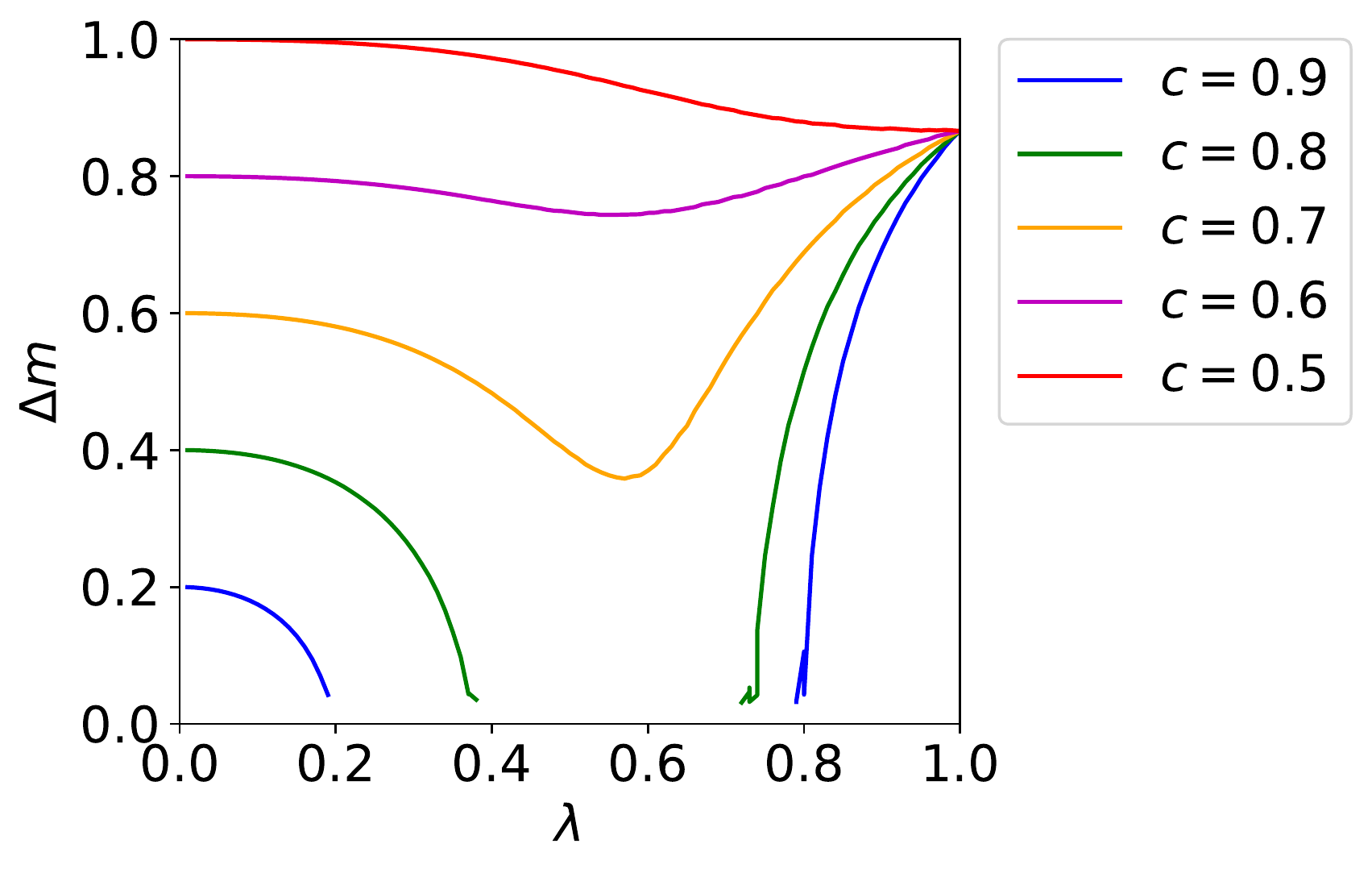}
          \hspace{1.6cm} (a)$\;p=3$
      \end{minipage}
      \begin{minipage}{0.5\linewidth}
        \begin{center}
          \includegraphics[width=8cm,clip]{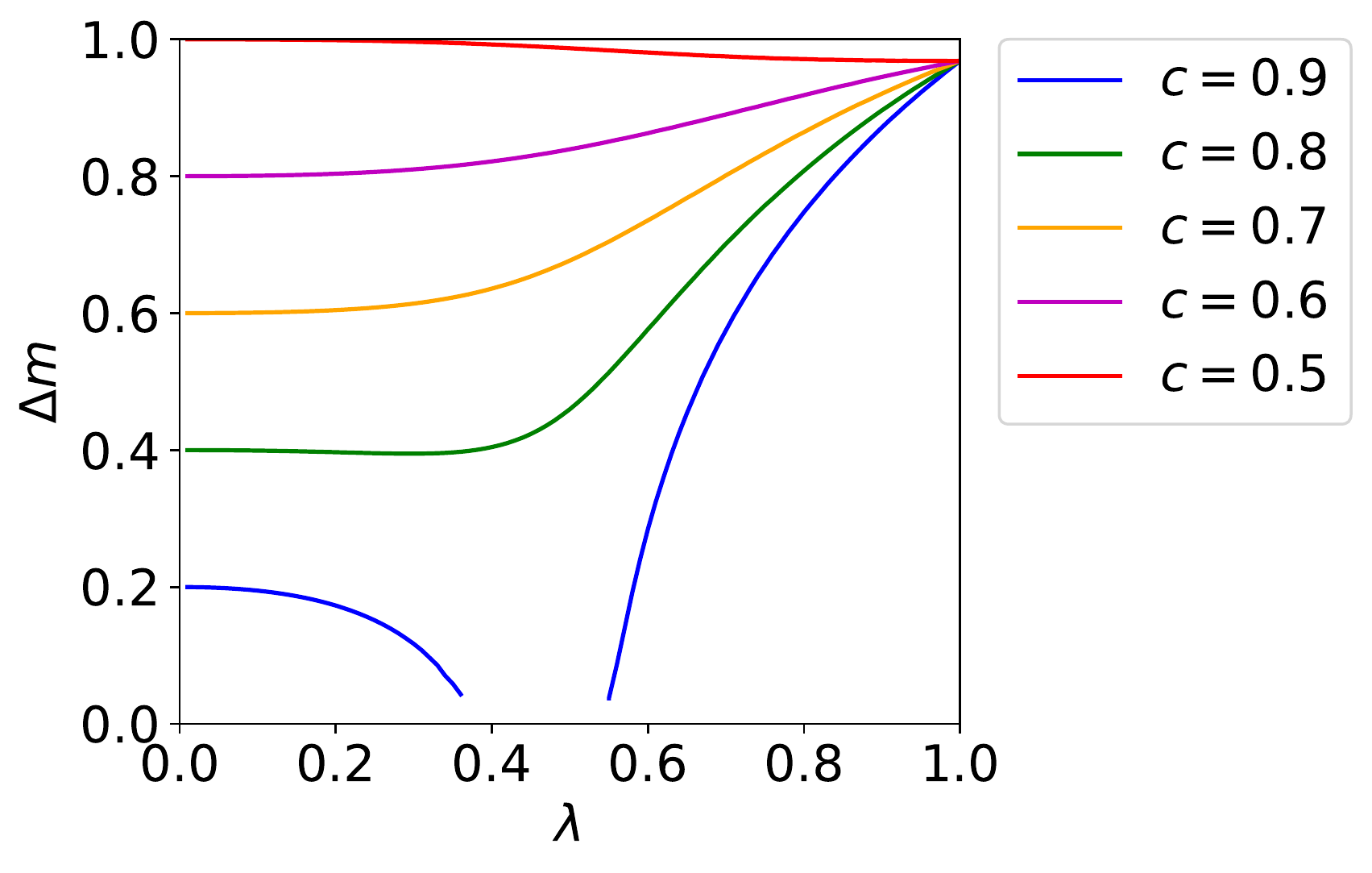}
          \hspace{1.6cm} (b)$\;p=5$
        \end{center}
      \end{minipage}
    \end{tabular}
    \caption{(Color online) Jump in magnetization along the first-order transition line for $p=3$ and 5 and several values of $c$.}
\label{fig:delta_m_NoRandom}
\end{figure*}

An interesting question is whether or not the first-order transition becomes weaker in some sense by reverse annealing even when the system is driven across the line of first-order transitions. Since the rate of quantum tunneling between two local minima in the energy landscape depends on the distance between the minima, it is interesting to see how the jump in magnetization at first-order transitions is affected by reverse annealing, because the jump magnitude is expected to be a proxy of the distance between two minima. Figure~\ref{fig:delta_m_NoRandom} shows the jump in magnetization $\Delta m$ along the line of first-order phase transitions. Indeed, the jump magnitude decreases with increasing $c$, and vanishes for $c$ greater than the critical value given in Table~\ref{table:p_c_value_noRandom}. The interval of vanishing $\Delta m$ corresponds to the break in the first-order transition line [compare, e.g., the $p=3$ and $c=0.8$ case with Fig.~\ref{fig:PDNoRandomField}(c)].
In this sense of decreased barrier width, Fig.~\ref{fig:delta_m_NoRandom} clearly indicates that an initial state with a relatively large value of $c$ ($c\ge 0.6$ for both values of $p$) facilitates quantum tunneling for $\lambda <1$ even when the system is driven across a first-order transition. This indicates that reverse annealing, if started from a state reflecting a modest amount of information about the final ground state, has the potential to mitigate the difficulty of quantum annealing even when one cannot avoid first-order transitions.

\begin{figure*}[thb]
\centering
\includegraphics[width=6.5cm,clip]{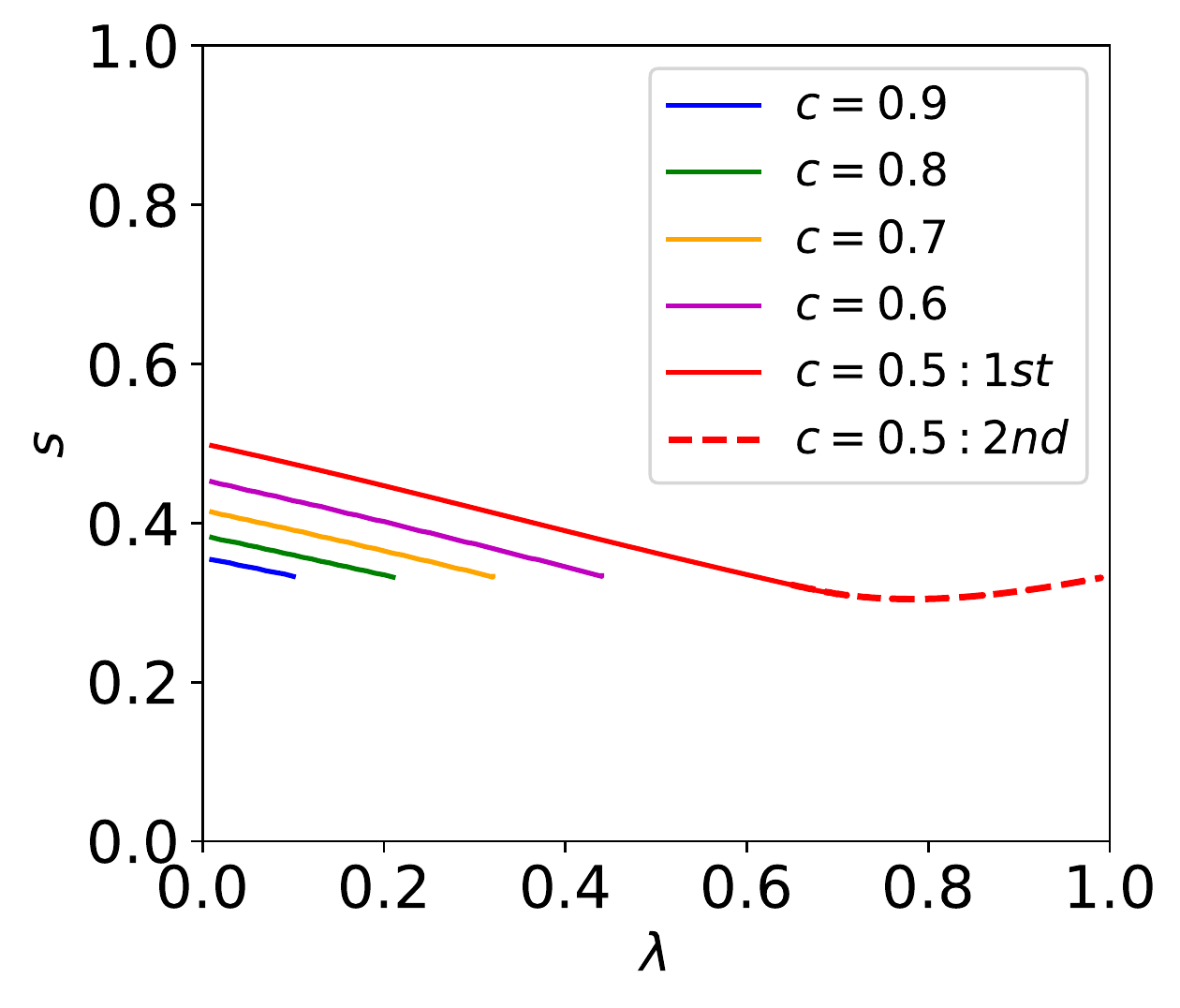}
\caption{(Color online) Phase diagram for $p=2$. Solid curves represent the line of first-order transitions, and the dashed curve is for second-order transition.}
\label{fig:souzu_p2_NoRandom}
\end{figure*}

A specific feature appears for $p=2$.  A shown in Fig.~\ref{fig:souzu_p2_NoRandom}, the second-order phase transition at $\lambda=1$ disappears immediately after $\lambda$ is reduced from $1$ except for the case of $c=0.5$. This is because the global inversion symmetry is broken in the initial state when $c>0.5$. For $c=0.5$, the second-order transition persists for $\lambda<1$ until it is replaced by a first-order transition.
\subsubsection{Random field with bimodal distribution}
\label{subsubsection:BinaryRandomField1}

We next study the case with random fields following the bimodal distribution,
\begin{align}
P(h_i)=\frac{1}{2}\delta(h_i-h_0)+\frac{1}{2}\delta(h_i+h_0) \quad (h_0 > 0).
\end{align}
The free energy and the self-consistent equation at zero temperature are
\begin{align}
\label{eq:free_energy_binary}
f=s(p-1)m^p&-\frac{c}{2}\sqrt{(spm^{p-1}+sh_0+(1-s)(1-\lambda))^2+(1-s)^2\lambda^2}\nonumber \\
&-\frac{1-c}{2}\sqrt{(spm^{p-1}+sh_0-(1-s)(1-\lambda))^2+(1-s)^2\lambda^2}\nonumber \\
&-\frac{c}{2}\sqrt{(spm^{p-1}-sh_0+(1-s)(1-\lambda))^2+(1-s)^2\lambda^2}\nonumber \\
&-\frac{1-c}{2}\sqrt{(spm^{p-1}-sh_0-(1-s)(1-\lambda))^2+(1-s)^2\lambda^2}
\end{align}
and
\begin{align}
\label{eq:self_consist_binary}
m=&\frac{c}{2}\frac{spm^{p-1}+sh_0+(1-s)(1-\lambda)}{\sqrt{(spm^{p-1}+sh_0+(1-s)(1-\lambda))^2+(1-s)^2\lambda^2}}\nonumber \\
&+\frac{1-c}{2}\frac{spm^{p-1}+sh_0-(1-s)(1-\lambda)}{\sqrt{(spm^{p-1}+sh_0-(1-s)(1-\lambda))^2+(1-s)^2\lambda^2}}\nonumber \\
&+\frac{c}{2}\frac{spm^{p-1}-sh_0+(1-s)(1-\lambda)}{\sqrt{(spm^{p-1}-sh_0+(1-s)(1-\lambda))^2+(1-s)^2\lambda^2}}\nonumber \\
&+\frac{1-c}{2}\frac{spm^{p-1}-sh_0-(1-s)(1-\lambda)}{\sqrt{(spm^{p-1}-sh_0-(1-s)(1-\lambda))^2+(1-s)^2\lambda^2}},
\end{align}
where $c$ is defined as before [Eq.~\eqref{eq:c}]. For the target Hamiltonian $s=\lambda=1$, the self-consistent equation (\ref{eq:self_consist_binary}) has two solutions, $m=0$ and $m=1$. The free energies of these solutions match $f(0)=f(1)$ at $h_0=1$. The state of the system is paramagnetic for $h_0>1$ and ferromagnetic for $h_0<1$. We focus our attention on the latter in the present paper.

\begin{figure*}[thb]
\centering
    \begin{tabular}{c}
      \begin{minipage}{0.5\linewidth}
			\centering
          \includegraphics[width=6cm,clip]{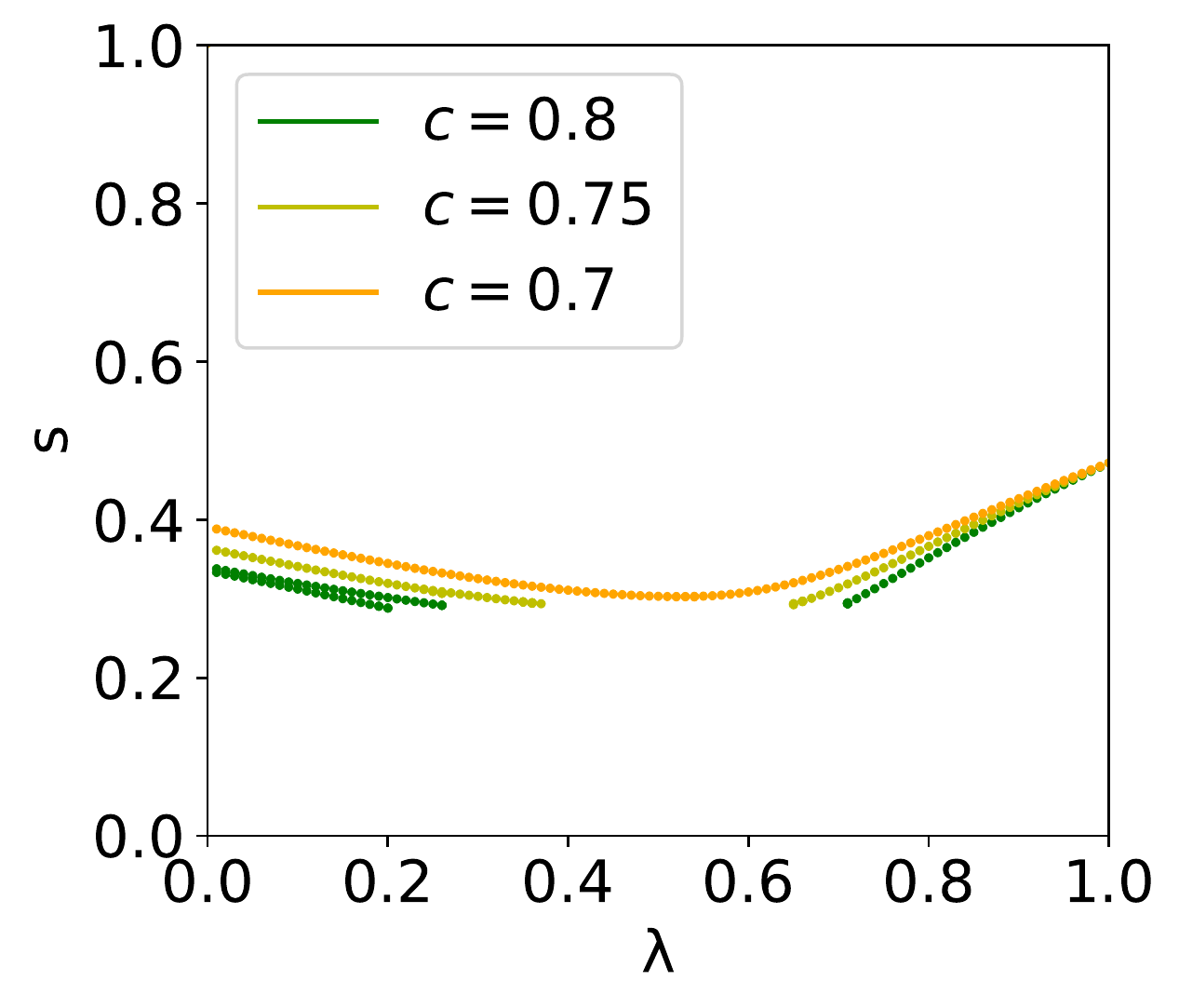}
          \hspace{1.6cm} (a)$\;p=3$
      \end{minipage}
	\begin{minipage}{0.5\linewidth}
			\centering
          \includegraphics[width=6cm,clip]{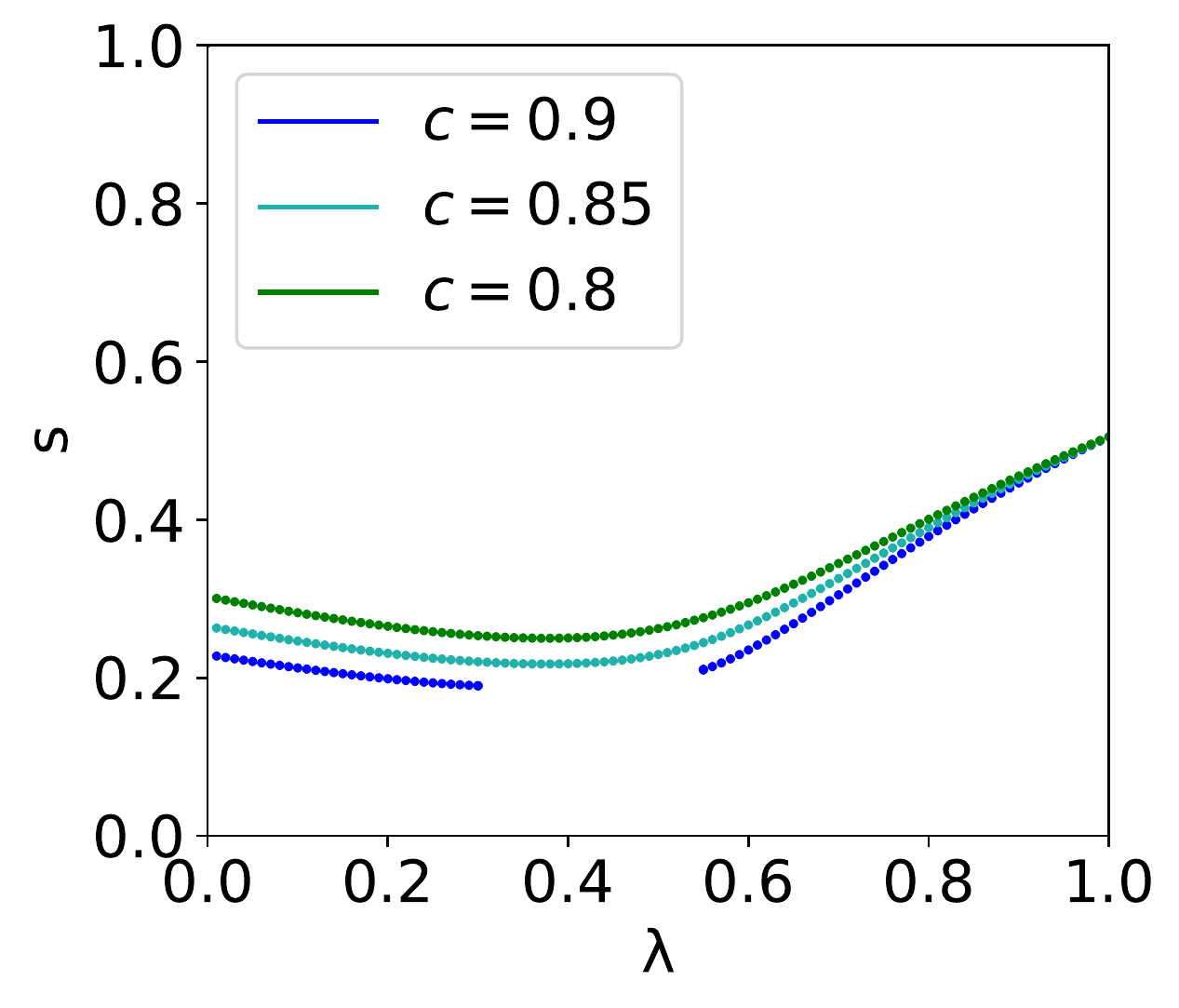}
          \hspace{1.6cm} (b)$\;p=5$
      \end{minipage}
    \end{tabular}
    \caption{(Color online)  Phase diagrams on the $s$-$\lambda$ plane for $p=3$ and $5$ under bimodal random field. Lines represent first-order phase transitions. The parameter $h_0$ for the amplitude of random field is 0.5. Each line is for first-order transitions.}
\label{fig:PD_random}
\end{figure*}

\begin{figure*}[thb]
\centering
    \begin{tabular}{c}
      \begin{minipage}{0.5\linewidth}
			\centering
          \includegraphics[width=8cm,clip]{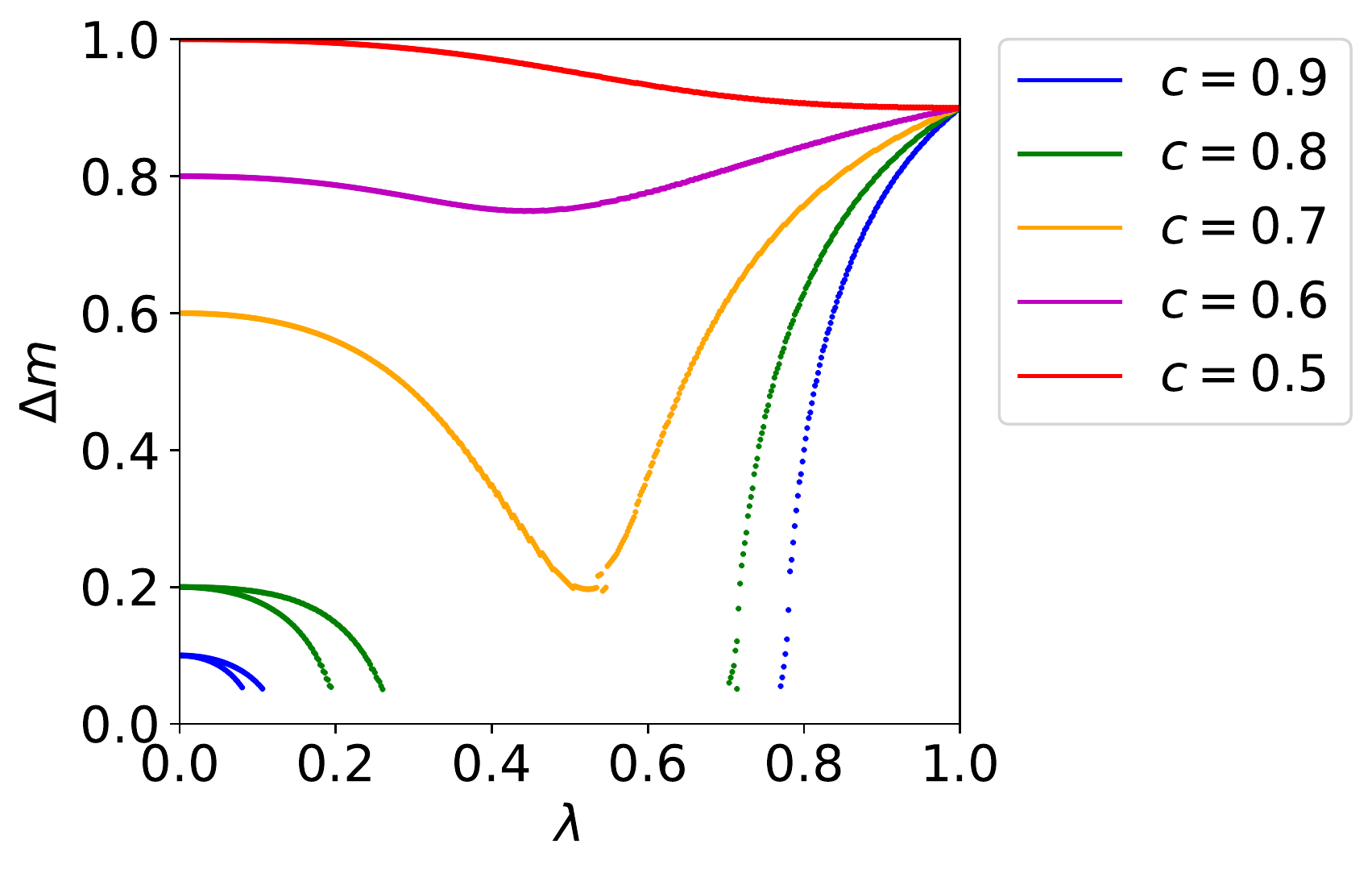}
          \hspace{1.6cm} (a)$\;p=3$
      \end{minipage}
      \begin{minipage}{0.5\linewidth}
        \begin{center}
          \includegraphics[width=8cm,clip]{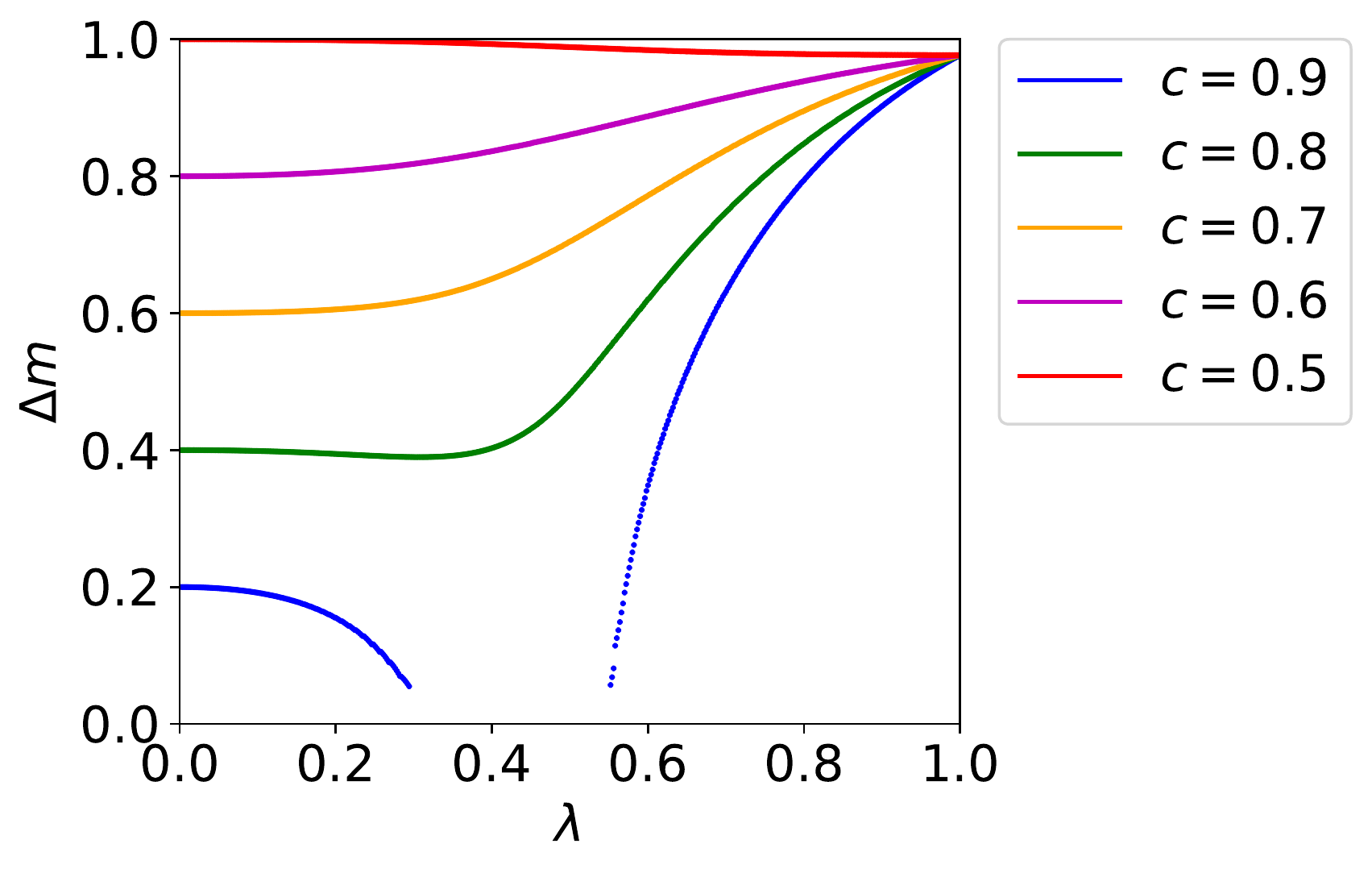}
          \hspace{1.6cm} (b)$\;p=5$
        \end{center}
      \end{minipage}
    \end{tabular}
    \caption{(Color online) Jump in magnetization at first-order transitions for $p=3$ and 5. The parameter $h_0$ is 0.5.}
\label{fig:delta_m_binary}
\end{figure*}

Phase diagrams are depicted in Fig.~\ref{fig:PD_random} for (a) $p=3$ and (b) $p=5$ in the case of $h_0=0.5$. Similarly to the previous case without random field, a larger value of $c$ leads to a wider break in the first-order transition line. A small difference from the case without random field is that there exist two transitions at and near $\lambda =0$ for a given value of $c$, as  can be seen by a careful inspection of Fig.~\ref{fig:PD_random} (a).  See Appendix \ref{appendix:B} for more details.
Table \ref{table:p_c_value_binary} lists critical values of $c$ where the first-order transition line starts to break into two parts.
\begin{table}[b]
	\caption{Critical values of $c$ where a break appears in the line of first-order transitions for the case with a bimodal random field.}
  \begin{tabular}{|c||c|c|c|c|c|c|c|c|c|c|c|c|} \hline
    $p$ &  \multicolumn{3}{c|}{3} & \multicolumn{3}{c|}{5} & \multicolumn{3}{c|}{7} & \multicolumn{3}{c|}{11} \\ \hline
    $h_0$ & 0.2 & 0.5 & 0.8 & 0.2 & 0.5 & 0.8 & 0.2 & 0.5 & 0.8 & 0.2 & 0.5 & 0.8 \\ \hline
	 $c$ & 0.74 & 0.72 & 0.7 & 0.89 & 0.89 & 0.87 & 0.94 & 0.93 & 0.93 & 0.97 & 0.97 & 0.97 \\ \hline
  \end{tabular}
	\label{table:p_c_value_binary}
\end{table}
Figure \ref{fig:delta_m_binary} is the jump in magnetization $\Delta m$ along the line of first-order transitions. Again the jump magnitude is smaller for $\lambda <1$ than for $\lambda =1$ if $c>0.5$, which may be interpreted to imply an increased tunneling rate by reverse annealing. 
\subsubsection{Gaussian random field}

\begin{figure*}[thb]
\centering
    \begin{tabular}{c}
      \begin{minipage}{0.5\linewidth}
			\centering
          \includegraphics[width=6cm,clip]{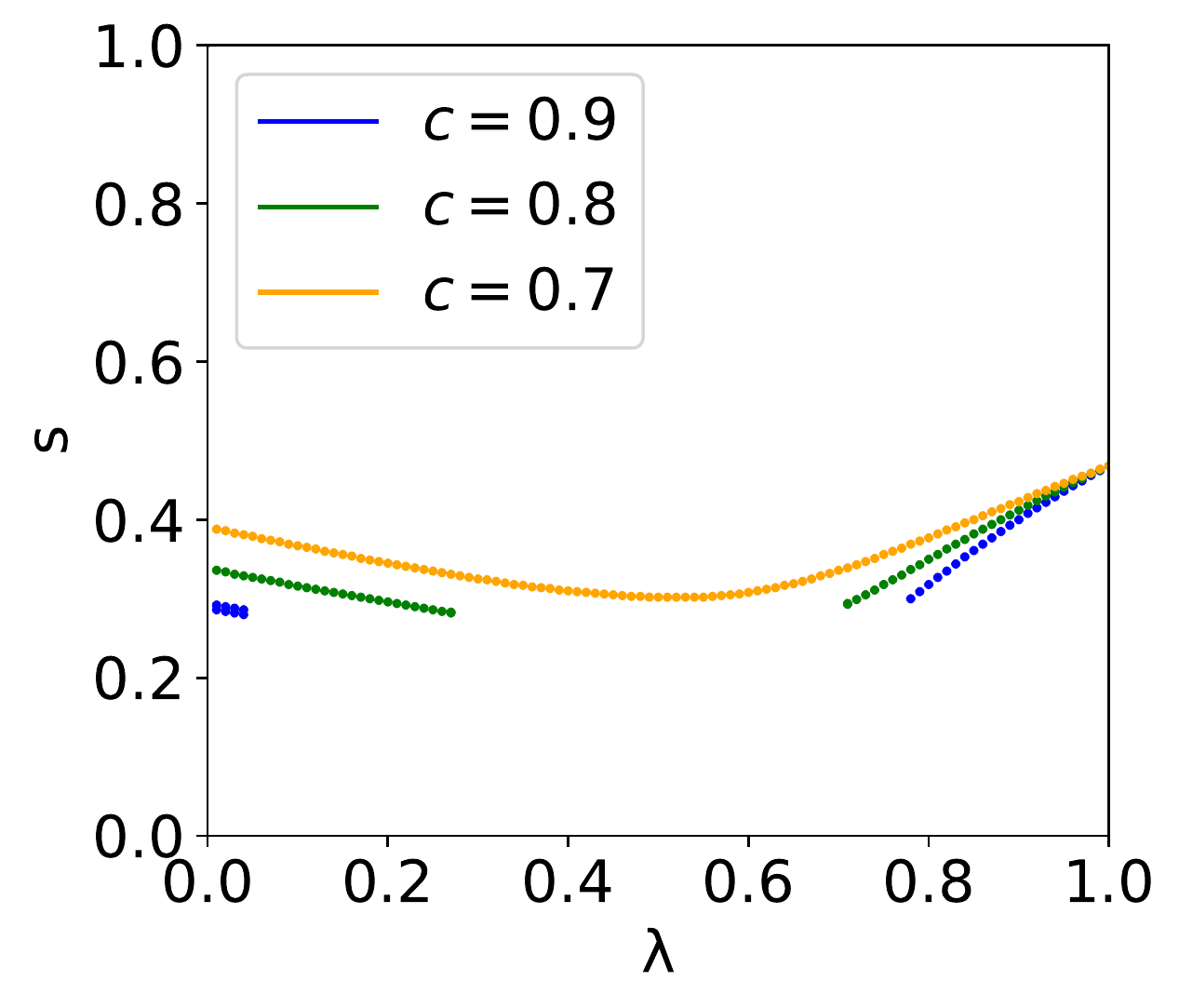}
          \hspace{1.6cm} (a)$\;p=3,\sigma=0.5$
      \end{minipage}
	\begin{minipage}{0.5\linewidth}
			\centering
          \includegraphics[width=6cm,clip]{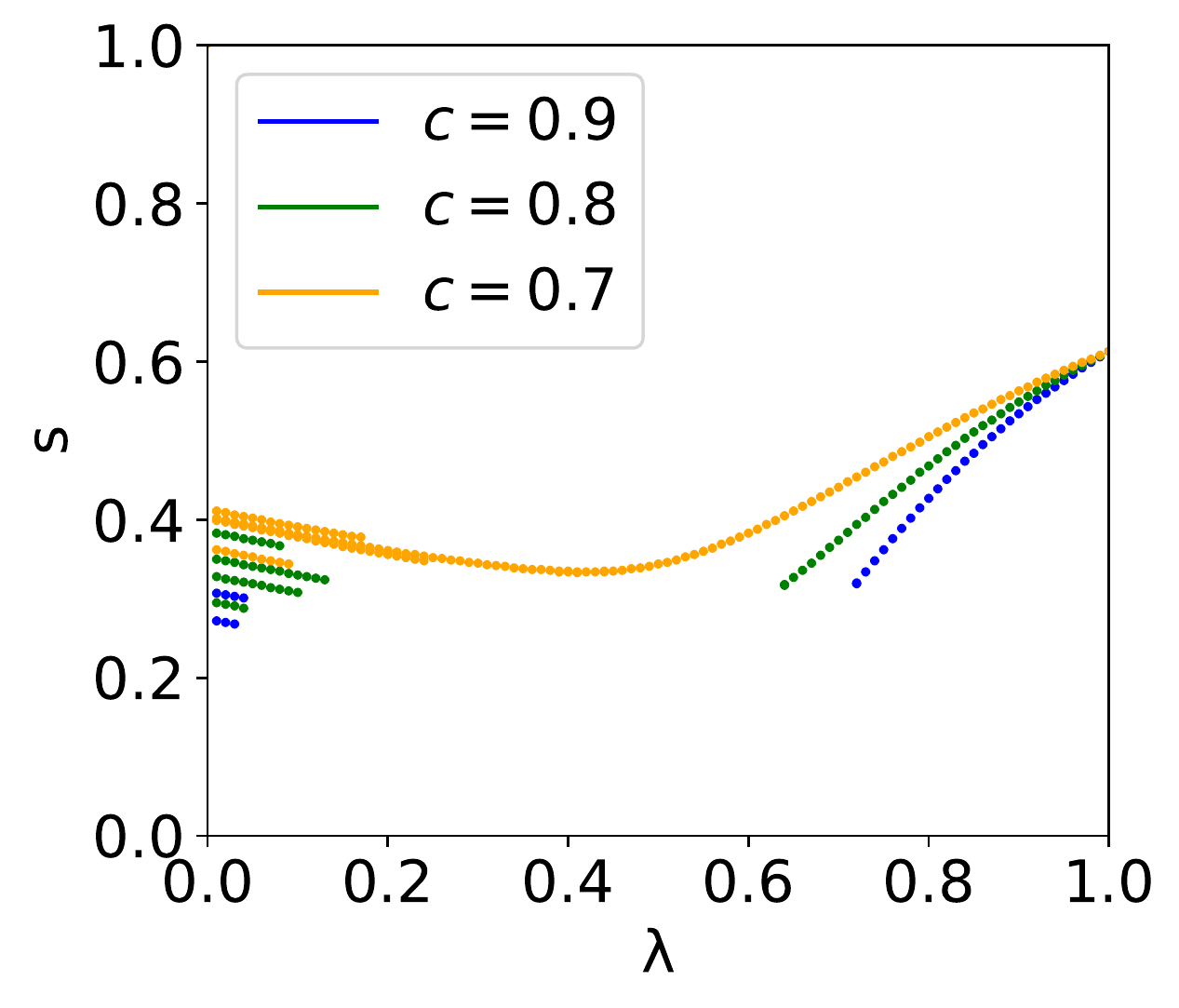}
          \hspace{1.6cm} (b)$\;p=3,\sigma=1$
      \end{minipage}
    \end{tabular}\\
    \begin{tabular}{c}
      \begin{minipage}{0.5\linewidth}
			\centering
          \includegraphics[width=6cm,clip]{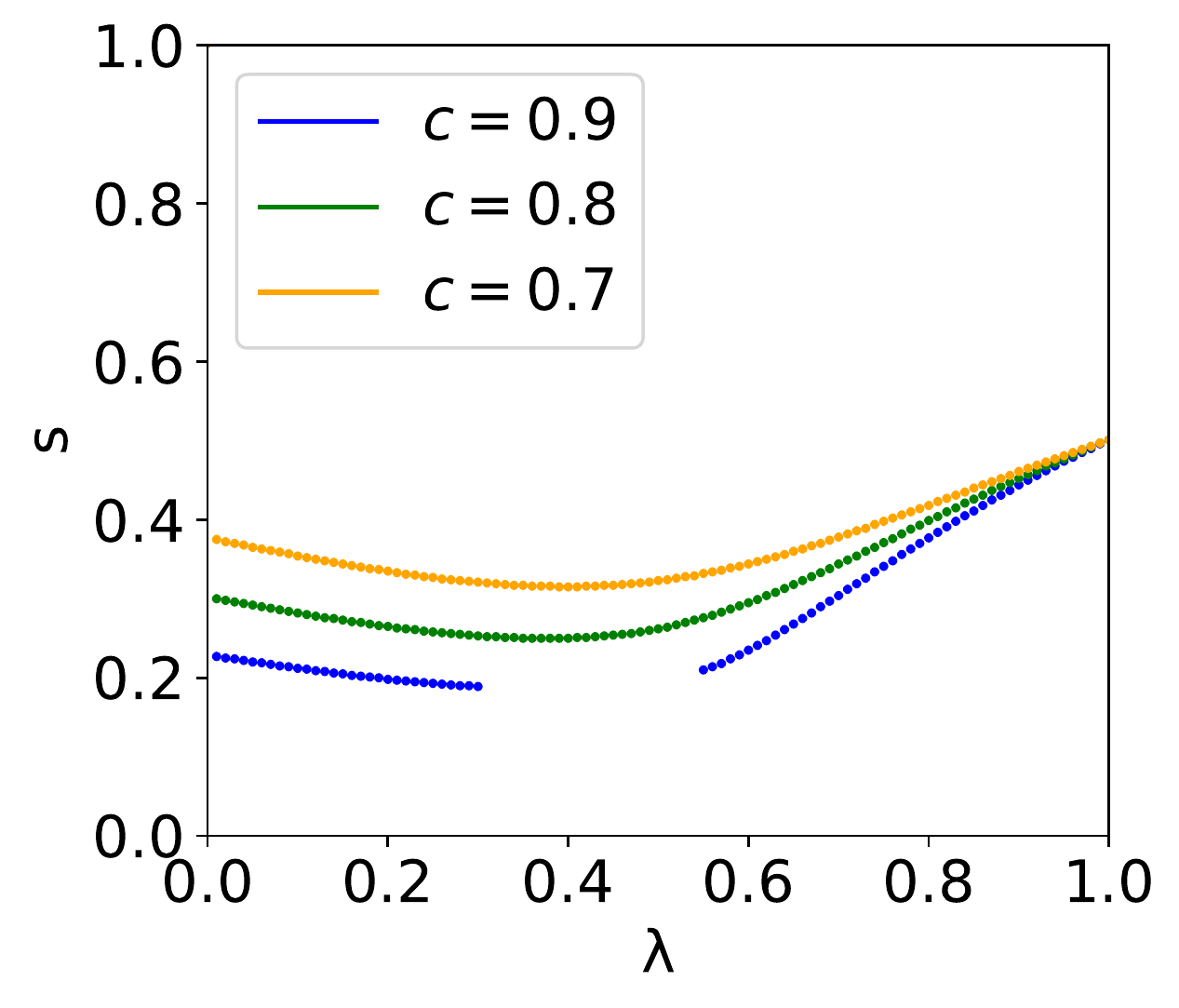}
          \hspace{1.6cm} (c)$\;p=5,\sigma=0.5$
      \end{minipage}
	\begin{minipage}{0.5\linewidth}
			\centering
          \includegraphics[width=6cm,clip]{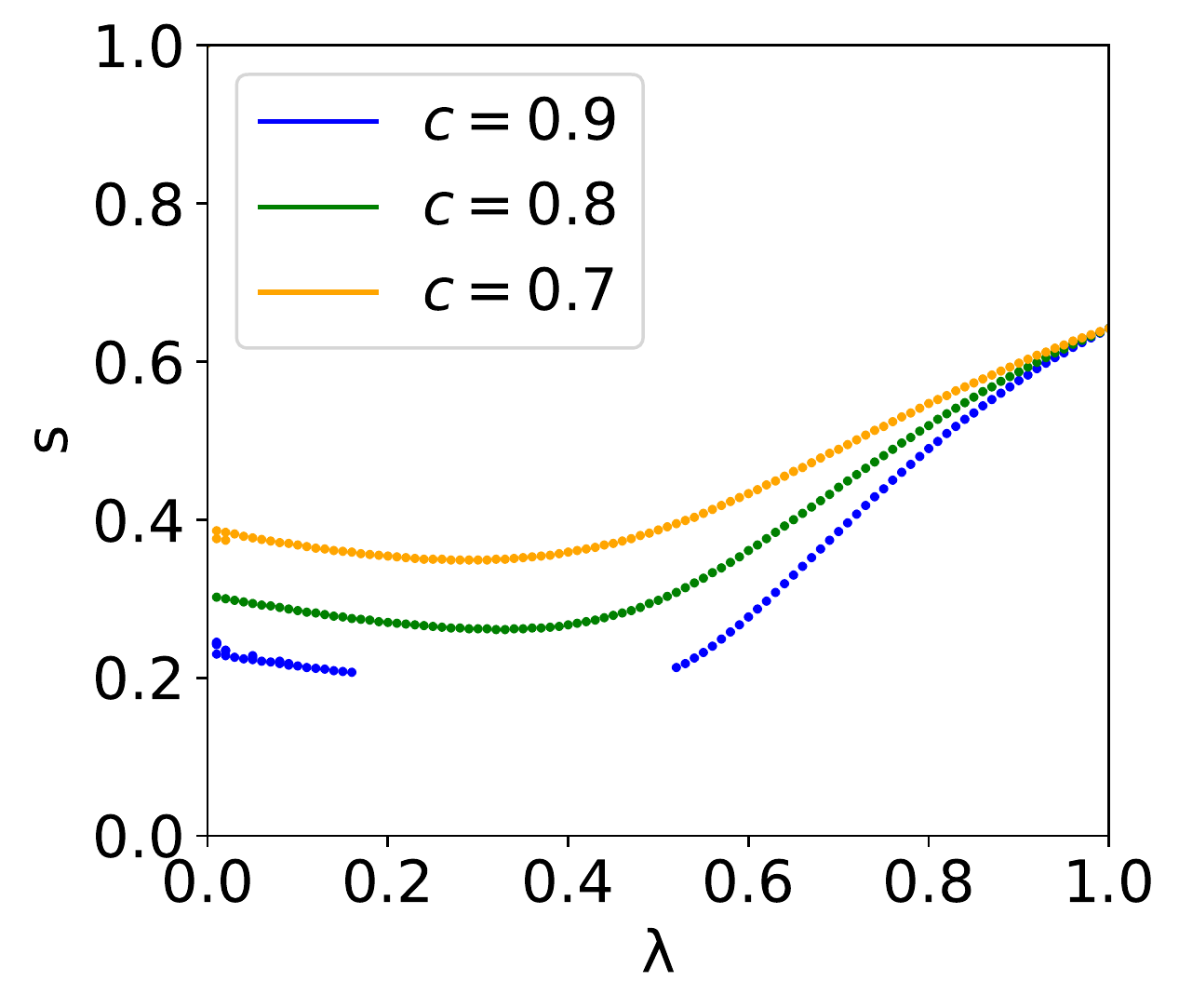}
          \hspace{1.6cm} (d)$\;p=5,\sigma=1$
      \end{minipage}
    \end{tabular}\\
    \caption{(Color online)  Phase diagrams on the $s$-$\lambda$ plane for (a) and (b) $p=3$, and (c) and (d) $p=5$  under the Gaussian-distributed random field. The standard deviation $\sigma$ is 0.5 and 1.}
\label{fig:PD_random_gaussian}
\end{figure*}

As the final example, we assume that the random field follows the Gaussian distribution,
\begin{align}
P(h)=\frac{1}{\sqrt{2\pi}\sigma}\exp \left(-\frac{h^2}{2\sigma^2}\right),
\end{align}
The free energy and the self-consistent equation are
\begin{align}
f=s(p-1)m^p&-\frac{c}{\sqrt{2\pi}\sigma}\int e^{-\frac{h^2}{2\sigma^2}}\sqrt{(spm^{p-1}+sh+(1-s)(1-\lambda))^2+(1-s)^2\lambda^2}dh \nonumber \\
&-\frac{1-c}{\sqrt{2\pi}\sigma}\int e^{-\frac{h^2}{2\sigma^2}}\sqrt{(spm^{p-1}+sh-(1-s)(1-\lambda))^2+(1-s)^2\lambda^2}dh
\end{align}
and
\begin{align}
\label{eq:self_consist_gaussian}
m=&\frac{c}{\sqrt{2\pi}\sigma}\int e^{-\frac{h^2}{2\sigma^2}}\frac{spm^{p-1}+sh+(1-s)(1-\lambda)}{\sqrt{(spm^{p-1}+sh+(1-s)(1-\lambda))^2+(1-s)^2\lambda^2}}dh \nonumber \\
&+\frac{1-c}{\sqrt{2\pi}\sigma}\int e^{-\frac{h^2}{2\sigma^2}}\frac{spm^{p-1}+sh-(1-s)(1-\lambda)}{\sqrt{(spm^{p-1}+sh-(1-s)(1-\lambda))^2+(1-s)^2\lambda^2}}dh.
\end{align}
Figure \ref{fig:PD_random_gaussian} shows the phase diagrams for $p=3$ and $5$ with $\sigma=0.5$ and $1$. 
When $s=\lambda=1$, the ferromagnetic solution of the self-consistent equation (\ref{eq:self_consist_gaussian}) has $m$ very close to $1$.  We can therefore regard $c$ as a good measure to gauge the proximity of the initial state to the correct solution, i.e., a larger $m$ means a closer initial state to the correct final state.  Figure \ref{fig:PD_random_gaussian}  again shows the existence of a break in the first-order line when $c$ is greater than a threshold. The main new feature compared to the bimodal case is the appearance of multiple first order transition lines near $\lambda=0$ when $\sigma=1$.

\begin{figure*}[thb]
\centering
    \begin{tabular}{c}
      \begin{minipage}{0.5\linewidth}
			\centering
          \includegraphics[width=8cm,clip]{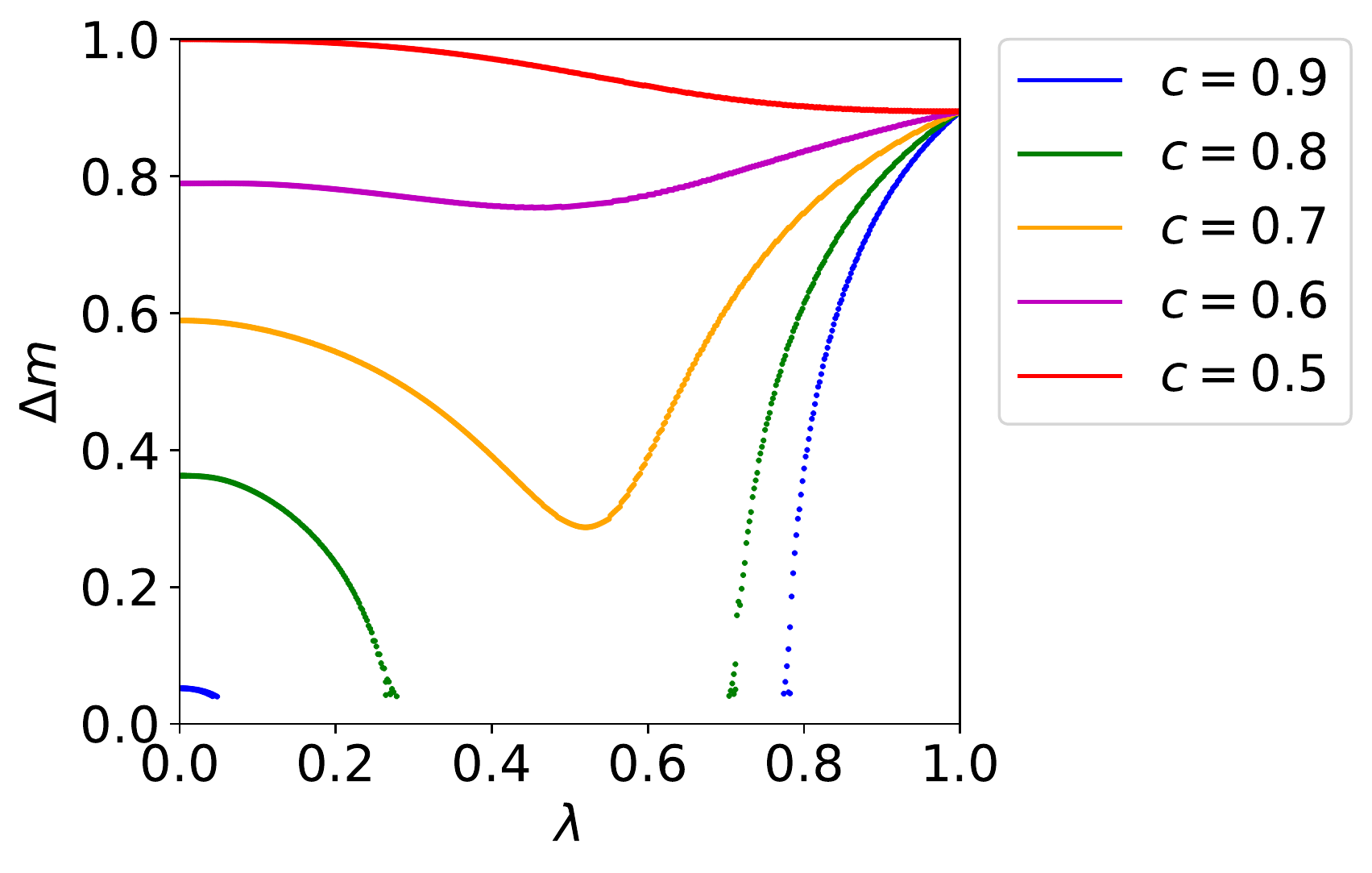}
          \hspace{1.6cm} (a)$\;p=3,\sigma=0.5$
      \end{minipage}
      \begin{minipage}{0.5\linewidth}
        \begin{center}
          \includegraphics[width=8cm,clip]{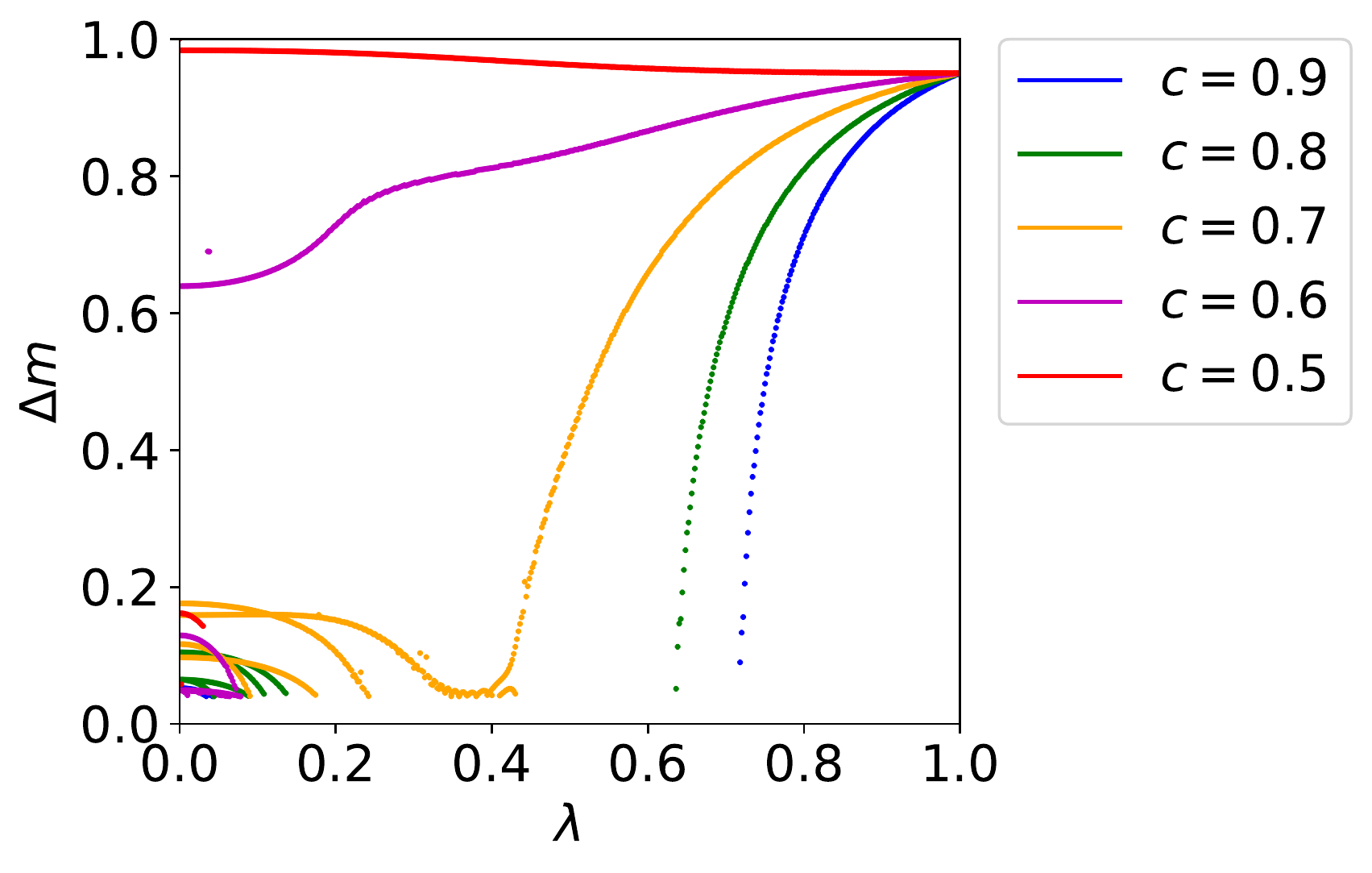}
          \hspace{1.6cm} (b)$\;p=3,\sigma=1$
        \end{center}
      \end{minipage}
    \end{tabular}
    \caption{(Color online) Jump in magnetization along the first-order transition line for $p=3$ under Gaussian random field.}
\label{fig:delta_m_gaussian}
\end{figure*}

In Table \ref{table:p_c_value_gaussian} we list critical values of $c$ when the first-order transition line starts to break up into two parts.
\begin{table}[htb]
	\caption{Critical values of $c$ for the first-order transition line for the Gaussian distribution random field.}
  \begin{tabular}{|c||c|c|c|c|c|c|c|c|} \hline
    $p$ &  \multicolumn{2}{c|}{3} & \multicolumn{2}{c|}{5} & \multicolumn{2}{c|}{7} & \multicolumn{2}{c|}{11} \\ \hline
    $\sigma$ & 0.5 & 1 & 0.5 & 1 & 0.5 & 1 & 0.5 & 1 \\ \hline
	 $c$ & 0.74 & 0.71 & 0.89 & 0.87 & 0.94 & 0.93 & 0.97 & 0.97 \\ \hline
  \end{tabular}
	\label{table:p_c_value_gaussian}
\end{table}
\newpage
Figure \ref{fig:delta_m_gaussian} is the jump in magnetization. We again observe that the jump is smaller for $\lambda<1$ than for $\lambda=1$ if the initial state is relatively close to the correct solution, signifying a potential increase in the tunneling rate. 
\section{Reverse Annealing for a non-stoquastic Hamiltonian}
\label{sec:non-stoquastic}

In this section, we formulate and solve the mean-field theory of reverse annealing for a non-stoquastic Hamiltonian.

The Hamiltonian of Eq.~(\ref{eq:hamiltonian1}) becomes non-stoquastic (has positive off-diagonal elements in the computational basis) \cite{Bravyi2008} by the introduction of antiferromagnetic transverse interactions \cite{Seki2012,Seoane2012,NishimoriTakada2017},
\begin{align}
\label{eq:hamiltonian_Nonst}
\hat{H}(s,\lambda)=s\nu\hat{H}_0+(1-s)(1-\lambda)\hat{H}_{{\rm init}}+(1-s)\lambda\hat{V}_{{\rm TF}}+(1-s)(1-\nu)\hat{V}_{{\rm AFTI}}\quad (0\le s, \lambda, \nu \le 1),
\end{align}
where
\begin{align}
\hat{V}_{{\rm AFTI}}=N\left(\frac{1}{N}\sum_{i=1}^N\hat{\sigma}_i^x\right)^2.
\end{align}
%
The new parameter $\nu$ controls the amplitude of the term $\hat{V}_{{\rm AFTI}}$. We note that the non-stoquasticity we consider here is ``curable"~\cite{Marvian:2018aa}, in the sense that it can be removed by a local unitary basis transformation: $\hat{\sigma}_i^x \leftrightarrow \hat{\sigma}_i^z$.
It is straightforward to solve this problem very similarly as before using the Suzuki-Trotter decomposition and the static approximation. The resulting free energy as a function of longitudinal and transverse magnetization, $m_z$ and $m_x$, is
\begin{align}
f=&(p-1)s\nu m_z^p-s(1-\nu)m_x^2\nonumber \\
&-T\left[\ln2\cosh\beta\sqrt{\left(ps\nu m_z^{p-1}+s\nu h_i+(1-s)(1-\lambda)\epsilon_i\right)^2+\left((1-s)\lambda-2s(1-\nu)m_x\right)^2}\right]_i
\end{align}
In the low-temperature limit $T\to 0$, the free energy and its minimization condition are
\begin{align}
f=&(p-1)s\nu m_z^p-s(1-\nu)m_x^2\nonumber \\
&-\left[\sqrt{\left(ps\nu m_z^{p-1}+s\nu h_i+(1-s)(1-\lambda)\epsilon_i\right)^2+\left((1-s)\lambda-2s(1-\nu)m_x\right)^2}\right]_i,
\end{align}
and
\begin{align}
m_z=\left[\frac{ps\nu m_z^{p-1}+s\nu h_i+(1-s)(1-\lambda)\epsilon_i}{\sqrt{\left(ps\nu m_z^{p-1}+s\nu h_i+(1-s)(1-\lambda)\epsilon_i\right)^2+\left((1-s)\lambda-2s(1-\nu)m_x\right)^2}}\right]_i,
\end{align}
\begin{align}
m_x=\left[\frac{(1-s)\lambda-2s(1-\nu)m_x}{\sqrt{\left(ps\nu m_z^{p-1}+s\nu h_i+(1-s)(1-\lambda)\epsilon_i\right)^2+\left((1-s)\lambda-2s(1-\nu)m_x\right)^2}}\right]_i.
\end{align}

\subsubsection{No random field}
\label{subsubsection:Non-sto_NoRandomField}
\begin{figure*}[thb] 
\centering
\includegraphics[width=10cm]{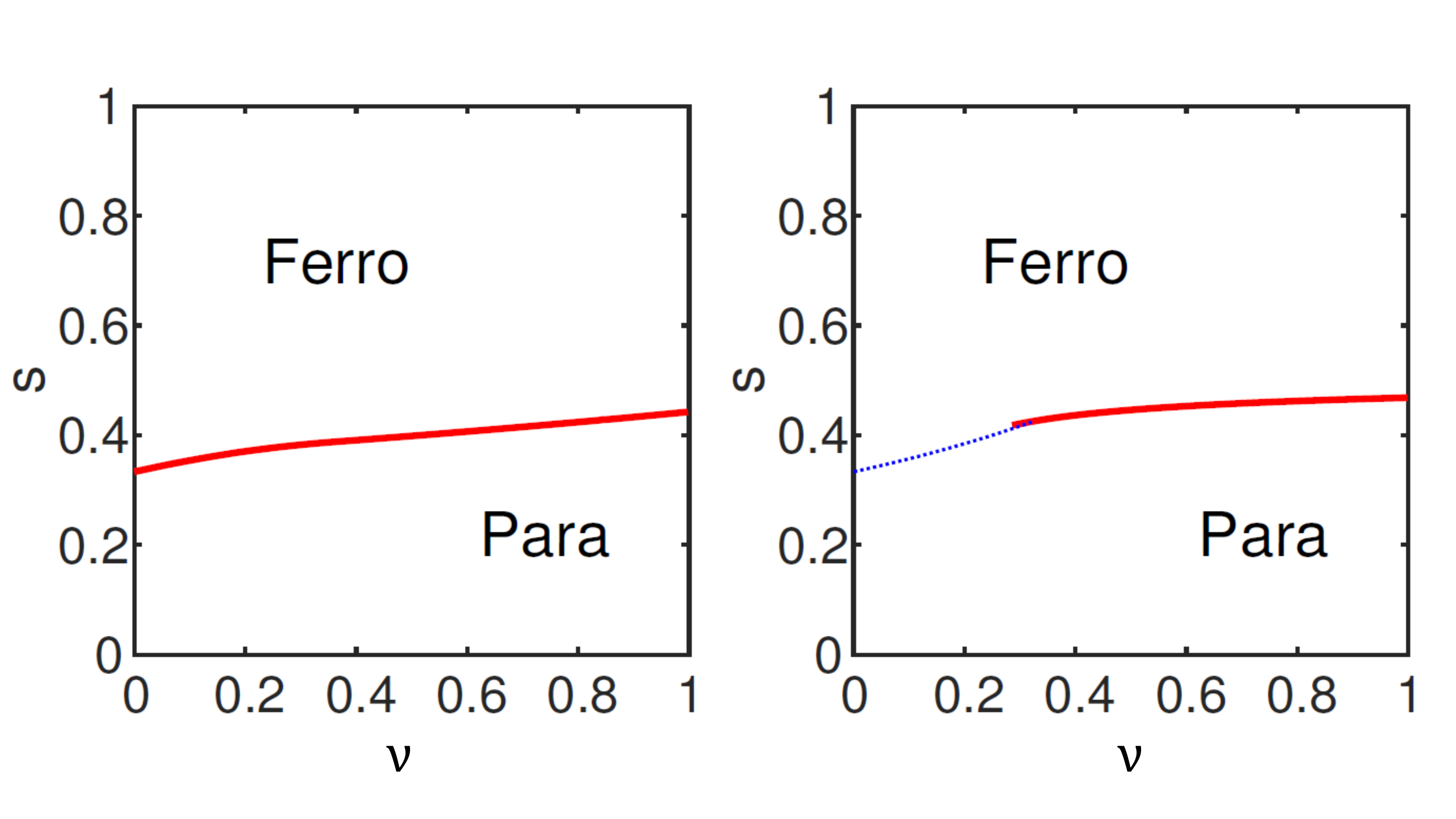}
\caption{(Color online) Phase diagram in the $s$-$\nu$ plane for $p=3$ (left) and $p=5$ (right) for  conventional quantum annealing $\lambda=1$ with a non-stoquastic Hamiltonian. The red curve is a line of first-order transitions between the paramagnetic and ferromagnetic phases. The blue dotted curve is for second-order transitions.}
\label{fig:Non-sto_s-vPhaseDiagram}
\end{figure*}

According to Ref.~\cite{Seki2012}, for conventional quantum annealing ($\lambda=1$) the first-order transition for $p>3$ can be avoided by the introduction of antiferromagnetic transverse interactions in the sense that first-order transitions are reduced to second order. Figure \ref{fig:Non-sto_s-vPhaseDiagram} is the phase diagram for $\lambda=1$ in the $s$-$\nu$ plane for $p=3$ (left) and $p=5$ (right). The case of $p=5$ has a line of second-order transitions for smaller $\nu$, shown by the dotted blue line, that replaces the first-order line for large $\nu$. This means that we can avoid first-order transitions by choosing an appropriate path in the phase diagram from the initial state at $s=0$ to the final state at $s=\nu=1$ in conventional quantum annealing.

\begin{figure*}[thb]
\centering
    \begin{tabular}{c}
      \begin{minipage}{0.33\linewidth}
			\centering
          \includegraphics[width=5cm,clip]{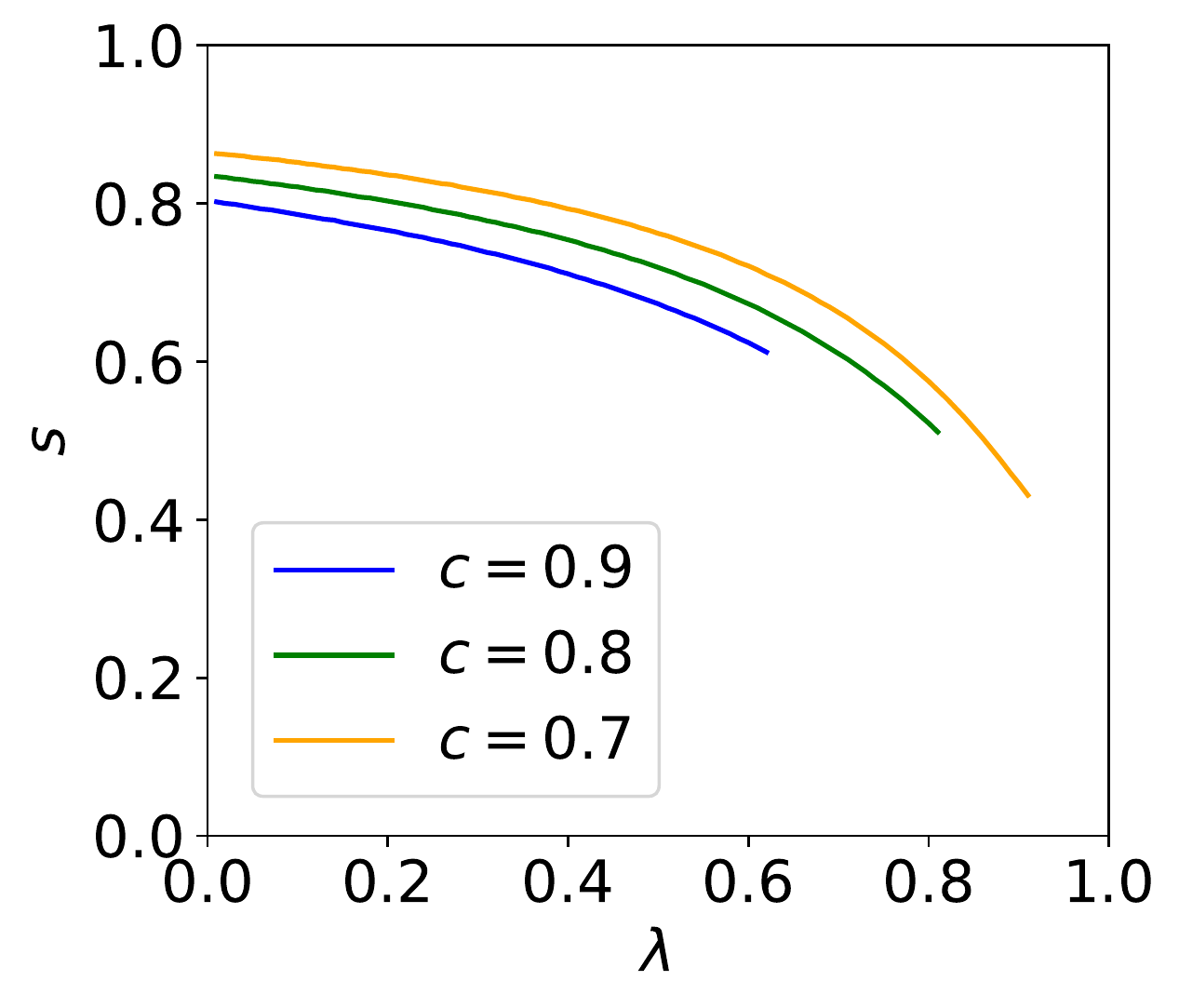}
          \hspace{1.6cm} (a)$\;p=3,\nu=0.1$
      \end{minipage}
      \begin{minipage}{0.33\linewidth}
        \begin{center}
          \includegraphics[width=5cm,clip]{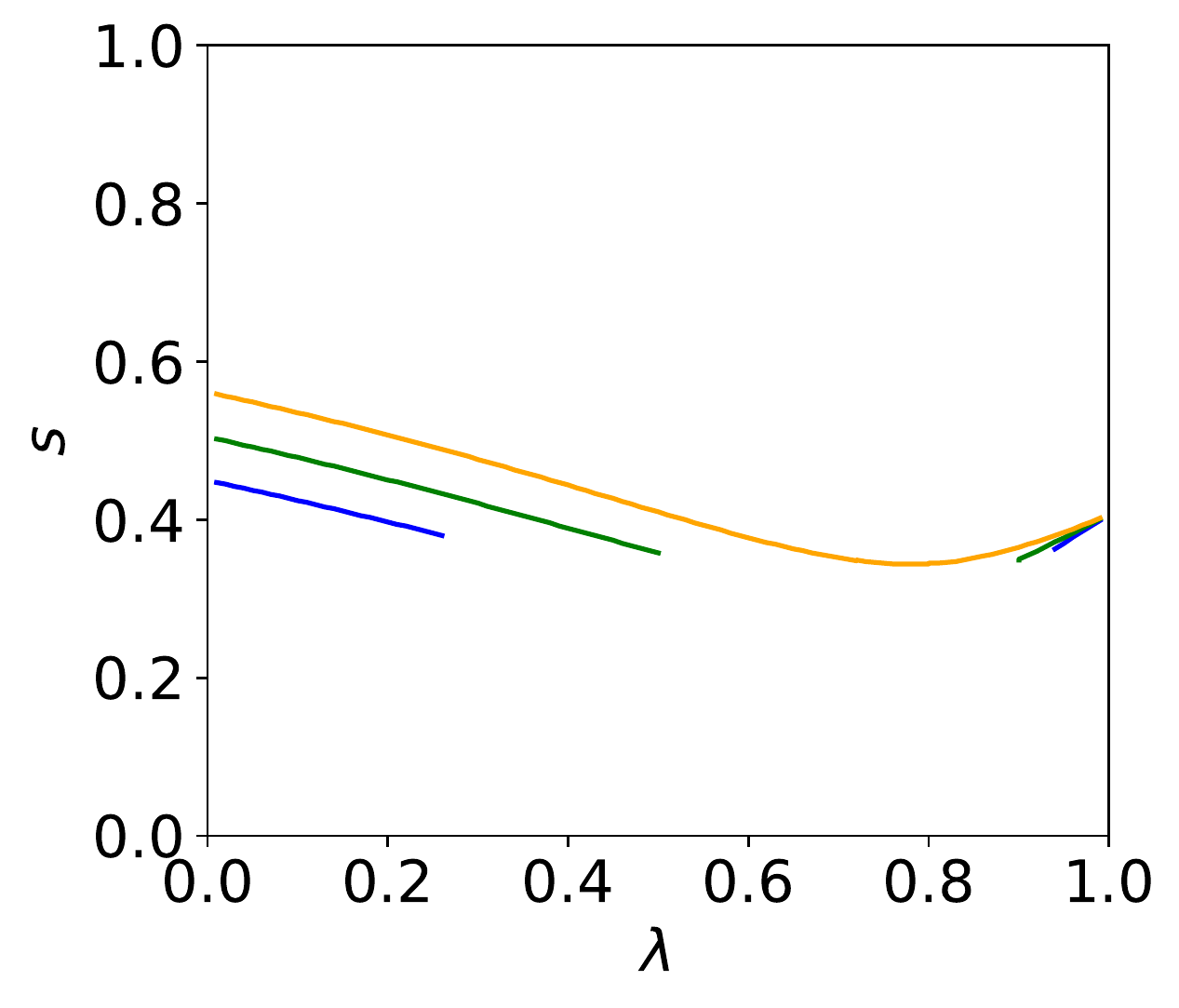}
          \hspace{1.6cm} (b)$\;p=3,\nu=0.5$
        \end{center}
      \end{minipage}
      \begin{minipage}{0.33\linewidth}
        \begin{center}
          \includegraphics[width=5cm,clip]{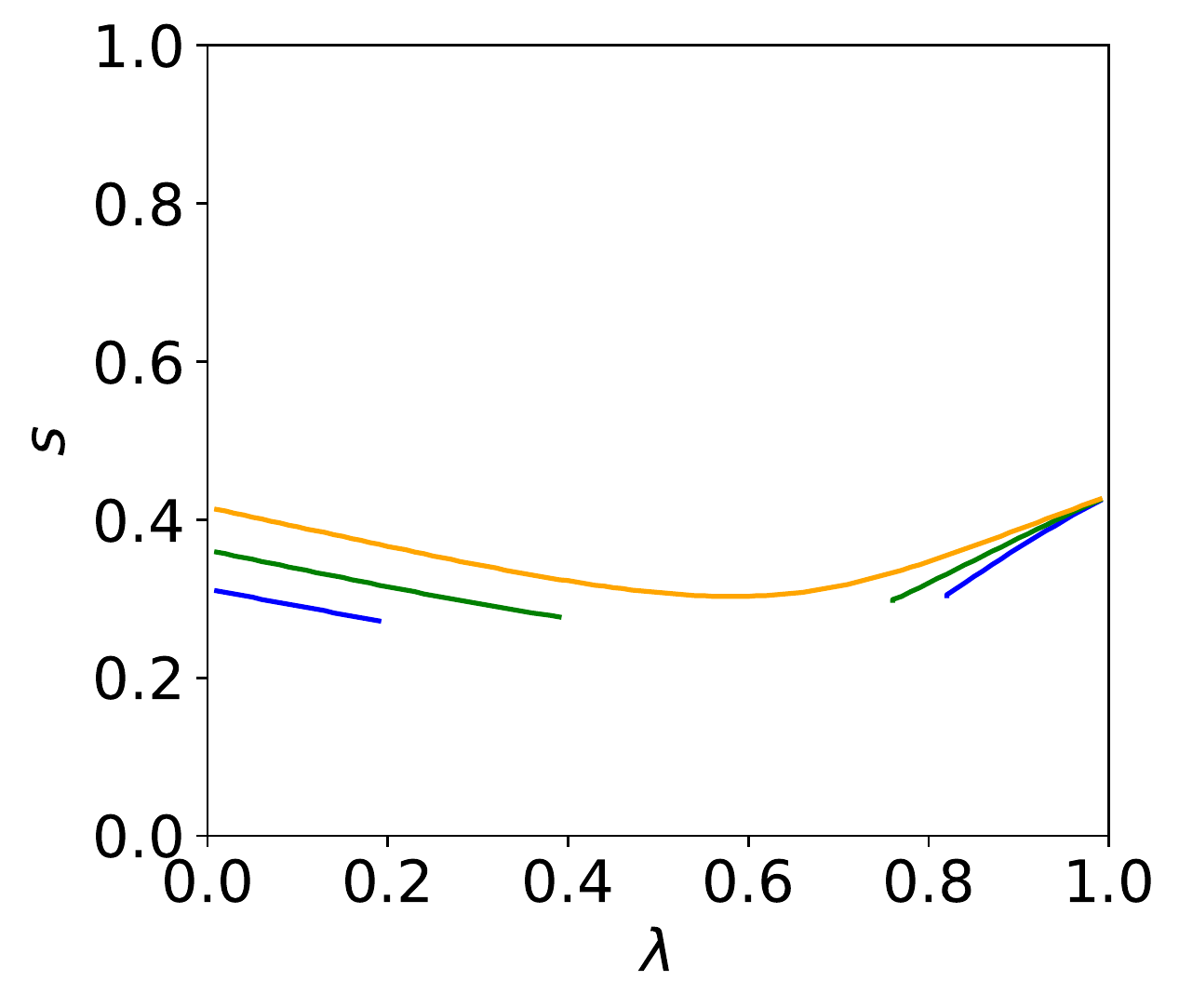}
          \hspace{1.6cm} (c)$\;p=3,\nu=0.9$
        \end{center}
      \end{minipage}\\

      \begin{minipage}{0.33\linewidth}
			\centering
          \includegraphics[width=5cm,clip]{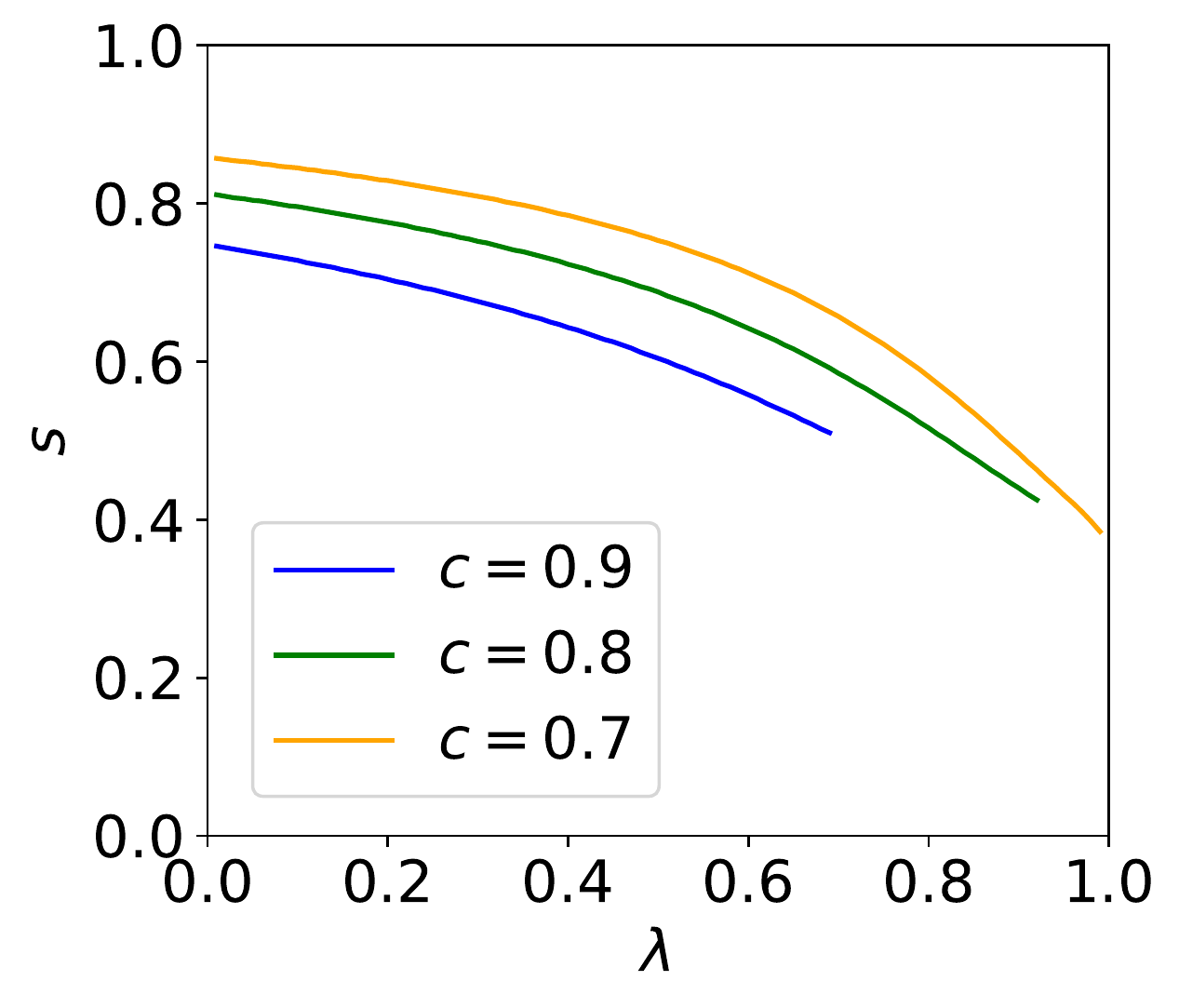}
          \hspace{1.6cm} (d)$\;p=5,\nu=0.1$
      \end{minipage}
      \begin{minipage}{0.33\linewidth}
        \begin{center}
          \includegraphics[width=5cm,clip]{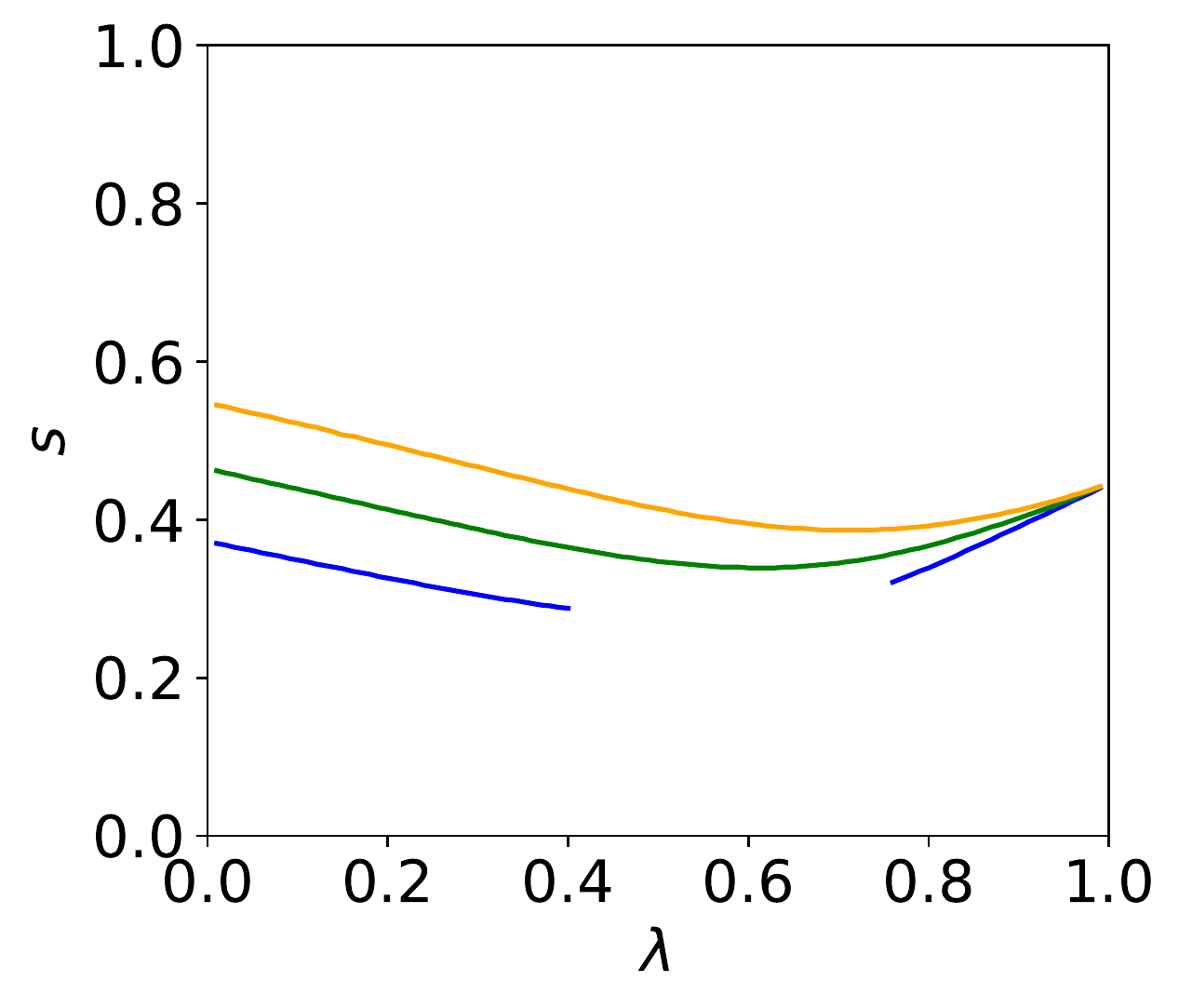}
          \hspace{1.6cm} (e)$\;p=5,\nu=0.5$
        \end{center}
      \end{minipage}
      \begin{minipage}{0.33\linewidth}
        \begin{center}
          \includegraphics[width=5cm,clip]{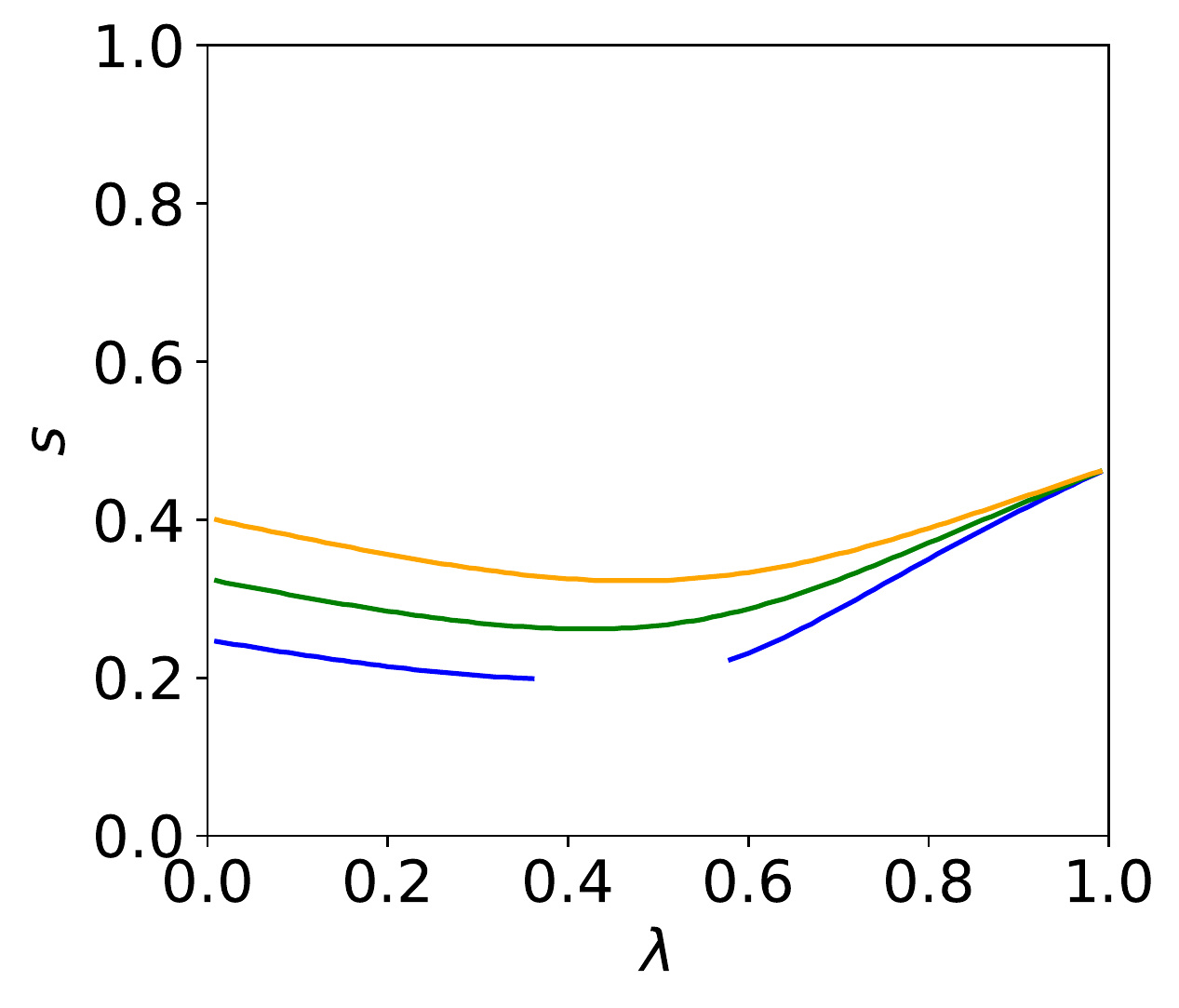}
          \hspace{1.6cm} (f)$\;p=5,\nu=0.9$
        \end{center}
      \end{minipage}
    \end{tabular}
    \caption{(Color online) Phase diagrams in the $s$-$\lambda$ plane for $p=3$ in (a)-(c), and $p=5$ in (d)-(f) for the non-stoquastic Hamiltonian. The parameter $\nu$ is set to $0.1, 0.5$, and $0.9$ (the smaller it is, the larger the amplitude of the non-stoquastic term). In panel (a), a first-order transition exists on the line $\lambda=1$, which disappears as soon as $\lambda$ becomes smaller than $1$ and then reappears for smaller $\lambda$. In panel (d), in contrast, the transition on the line $\lambda=1$ is of second order, as seen in Fig.~\ref{fig:Non-sto_s-vPhaseDiagram} (right), which is replaced by a line of first-order transitions for $\lambda$ below a threshold value.
    }
\label{fig:Non-sto_NoRandom_phasediagram}
\end{figure*}

Figure~\ref{fig:Non-sto_NoRandom_phasediagram} shows the phase diagrams in the $s$-$\lambda$ plane for $p=3$ and $p=5$ with $\nu=0.1, 0.5$, and $0.9$. In these phase diagrams, the transition points on the line of $\lambda=1$ coincide with the corresponding points in Fig.~\ref{fig:Non-sto_s-vPhaseDiagram}.  In panels (b), (c), (e), and (f), the line of first-order transitions is broken up into two parts for large $c$, similarly to the previous cases. Panel (a) has a first-order transition on the line $\lambda=1$, which disappears immediately after $\lambda$ is reduced from $1$, and reappears for smaller $\lambda$.  Panel (d) is similar except that the transition on the line $\lambda=1$ is of second order. 

These results suggest that reverse annealing works similarly for the non-stoquastic case, except when the amplitude of the term $\hat{V}_{\rm AFTI}$ that makes the Hamiltonian non-stoquastic is large, as in panels (a) and (d) of Fig.~\ref{fig:Non-sto_NoRandom_phasediagram}. In this case, rather than creating a break in the first order phase transition lines at intermediate values of the reverse annealing parameter $\lambda$, the break occurs for $\lambda$ larger than a threshold value. Since $\lambda=1$ corresponds to conventional quantum annealing, this can be interpreted as the non-stoquasticity favoring conventional over reverse quantum annealing.

\subsubsection{Random field with bimodal distribution}
\label{subsubsection:Non-sto_Binary}

\begin{figure*}[thb]
\centering
    \begin{tabular}{c}
      \begin{minipage}{0.33\linewidth}
			\centering
          \includegraphics[width=5cm,clip]{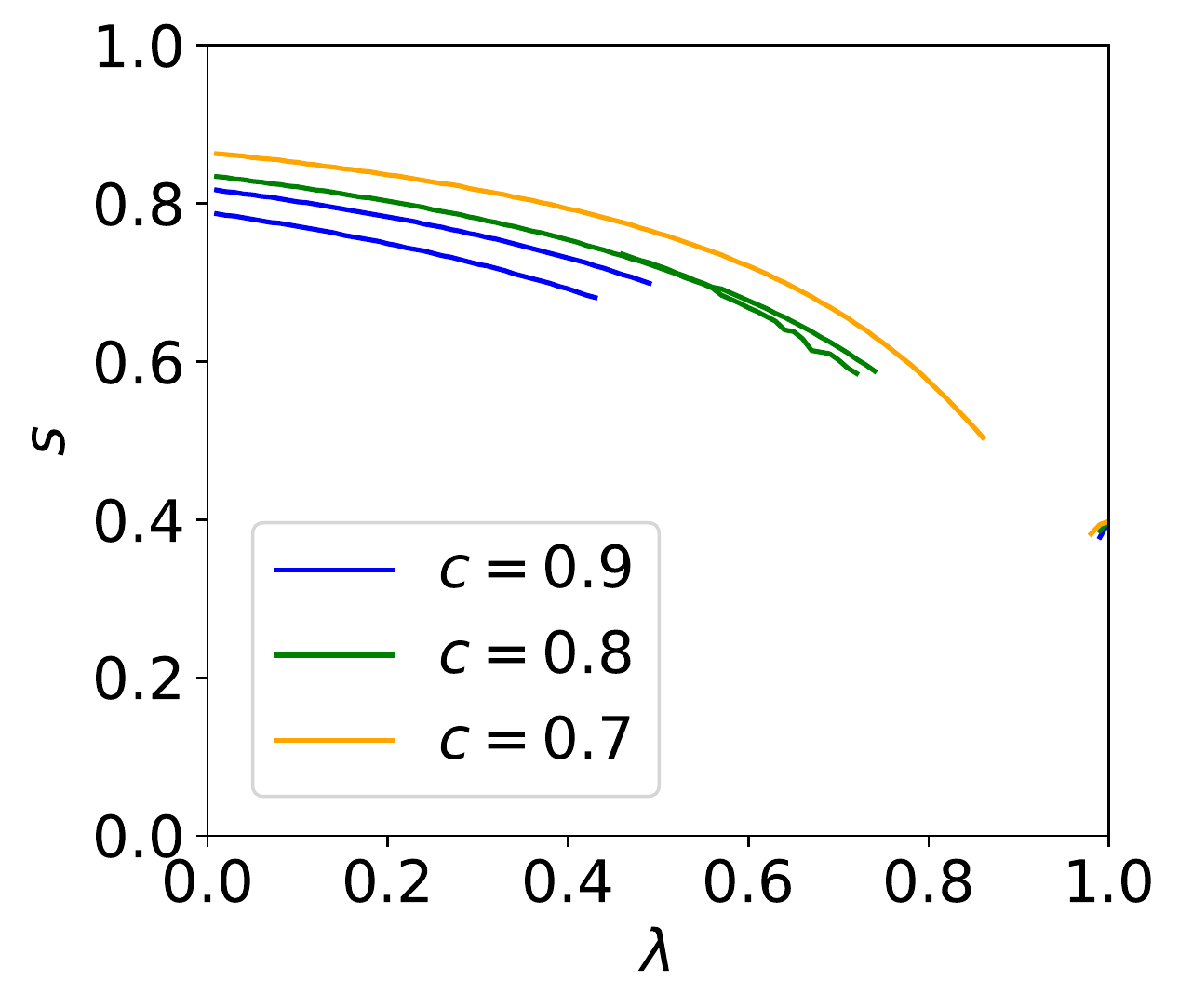}
          \hspace{1.6cm} (a)$\;p=3,\nu=0.1$
      \end{minipage}
      \begin{minipage}{0.33\linewidth}
        \begin{center}
          \includegraphics[width=5cm,clip]{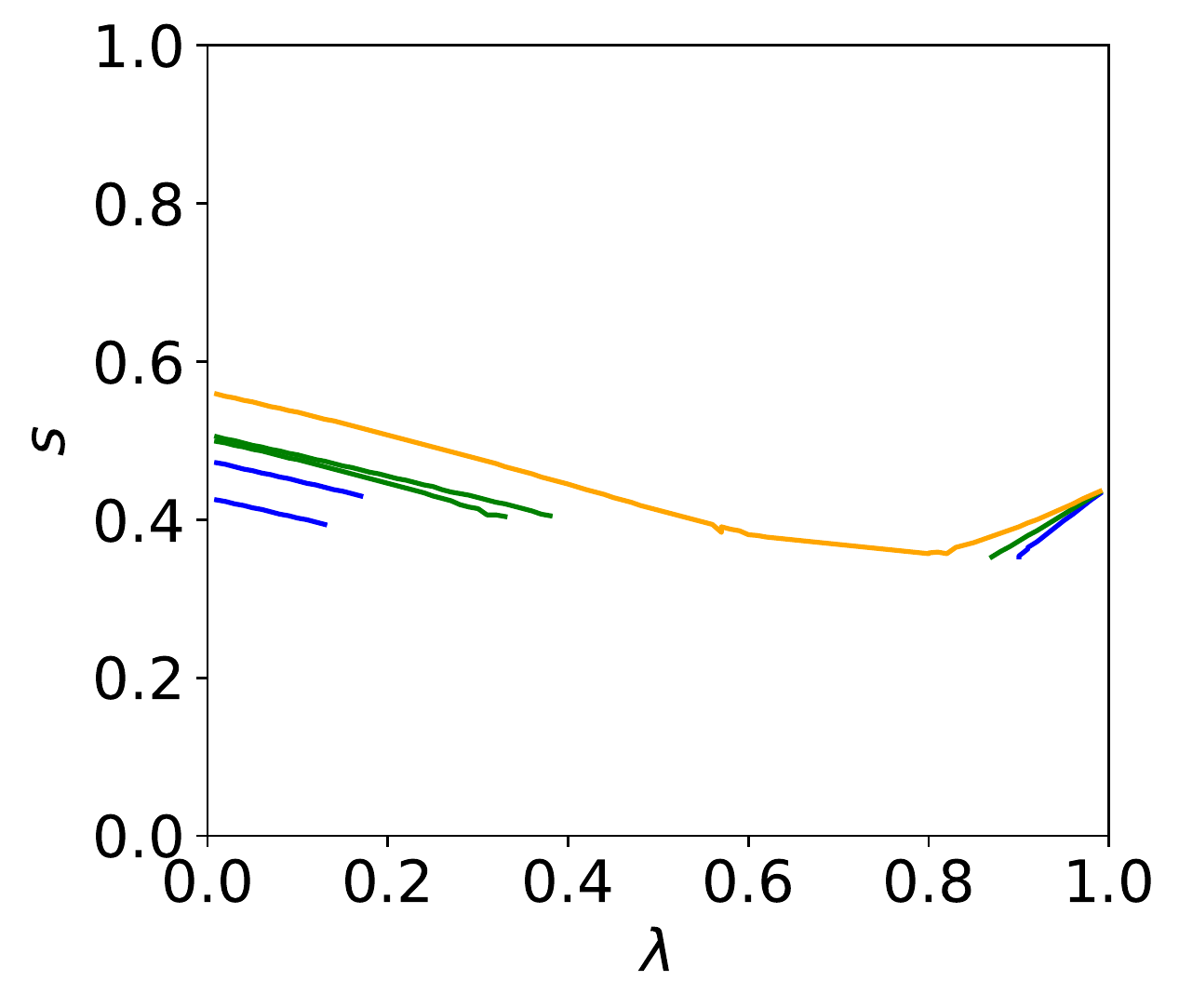}
          \hspace{1.6cm} (b)$\;p=3,\nu=0.5$
        \end{center}
      \end{minipage}
      \begin{minipage}{0.33\linewidth}
        \begin{center}
          \includegraphics[width=5cm,clip]{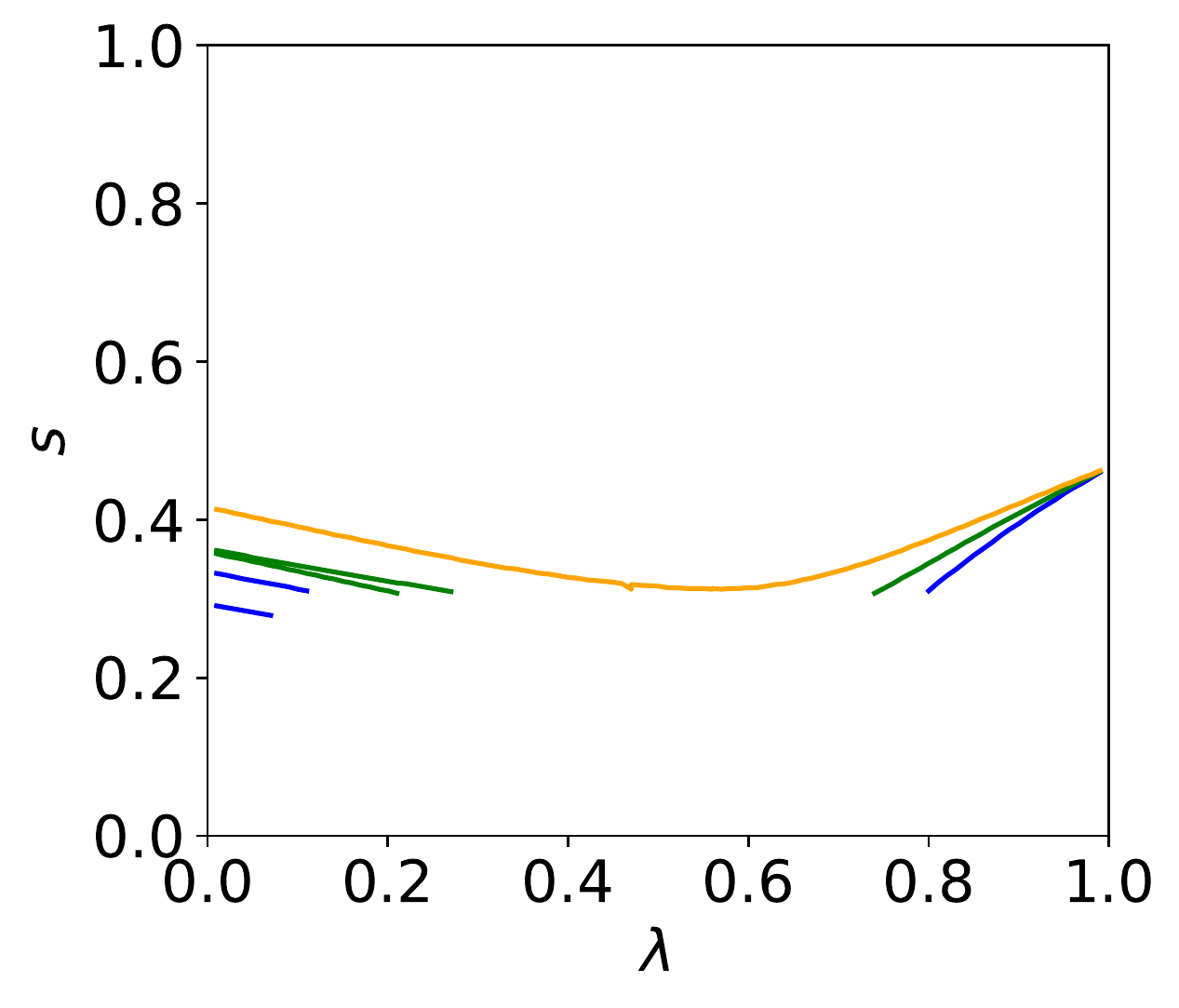}
          \hspace{1.6cm} (c)$\;p=3,\nu=0.9$
        \end{center}
      \end{minipage}\\

      \begin{minipage}{0.33\linewidth}
			\centering
          \includegraphics[width=5cm,clip]{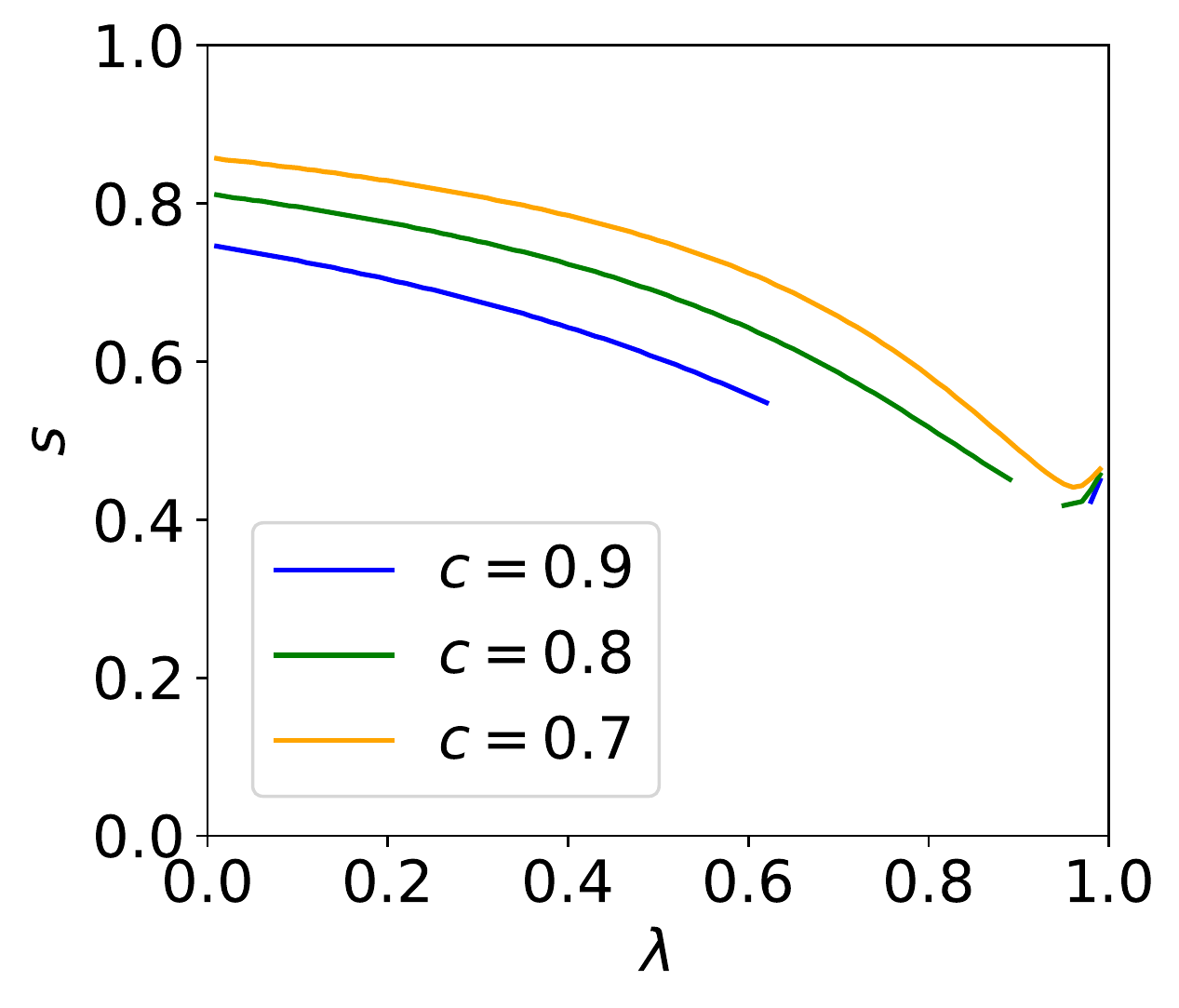}
          \hspace{1.6cm} (d)$\;p=5,\nu=0.1$
      \end{minipage}
      \begin{minipage}{0.33\linewidth}
        \begin{center}
          \includegraphics[width=5cm,clip]{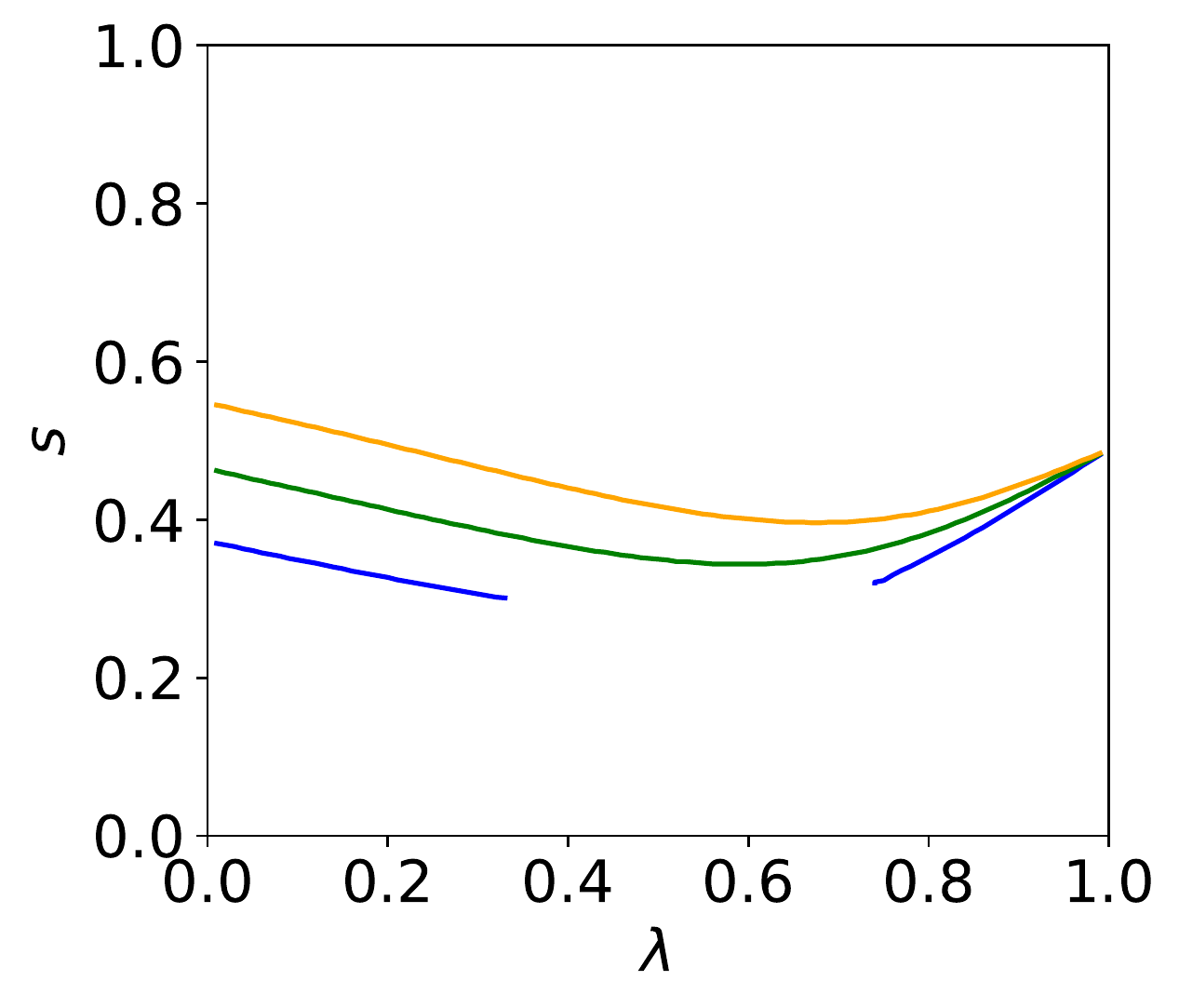}
          \hspace{1.6cm} (e)$\;p=5,\nu=0.5$
        \end{center}
      \end{minipage}
      \begin{minipage}{0.33\linewidth}
        \begin{center}
          \includegraphics[width=5cm,clip]{Non-sto_souzu_binary_p5_n05_h05.pdf}
          \hspace{1.6cm} (f)$\;p=5,\nu=0.9$
        \end{center}
      \end{minipage}
    \end{tabular}
    \caption{(Color online) Phase diagrams in the $s$-$\lambda$ plane for the case of bimodal random field for $p=3$ in panels (a)-(c), and $p=5$ in panels (d)-(f). The parameter $\nu$ is $0.1, 0.5$, and $0.9$, and $h_0=0.5$ everywhere.}
\label{fig:Non-sto_Binary_phasediagram}
\end{figure*}

We carried out a similar analysis for the case with bimodal random fields. The results are depicted in Fig.~\ref{fig:Non-sto_Binary_phasediagram}, where the amplitude of the random field is chosen to be $h_0=0.5$. The qualitative behavior remains the same as in the case without random field.

\section{Conclusion}
\label{sec:conclusions}
We have formulated and solved a mean-field theory of reverse annealing for the $p$-spin ferromagnetic model with and without longitudinal random fields as a problem of equilibrium statistical mechanics.  The results show that a path exists connecting the initial and final states in the phase diagram, along which there is no phase transition, if the initial state is close to the correct final ground state. Since the $p$-spin ferromagnetic model has a trivial solution (all spins up or down) that can be found by inspection, this indicates that the difficulties experienced by conventional quantum annealing are in this sense an artifact that can be ameliorated using reverse annealing. Of course, this begs the question of whether reverse annealing can also be useful for hard optimization problems. One indication that this might be so is that even when the annealing process goes across a first-order transition by an inappropriate choice of a path in the phase diagram or due to an inappropriate initial state, we have found that the jump in magnetization at a first-order transition is smaller than in the conventional method, provided that the initial state is not too far away from the correct final state. Since the quantum tunneling rate between two local minima in the free energy landscape is larger if the distance between the minima is smaller, we expect that reverse annealing can quantitatively enhance the performance of quantum annealing in many cases.

It should be stressed that the analysis presented here is of a purely static nature. Therefore, while it is tempting to conclude that avoidance of a first order phase transitions means an exponential speedup as compared to conventional quantum annealing as long as the system follows an adiabatic time evolution, caution must be exercised since our analysis does not include dynamics in any sense.
At vanishing temperature, a link to dynamics is available through the adiabatic theorem via the behavior of the energy gap $\Delta$ as a function of the system size $N$, e.g., $\Delta \propto e^{-aN}$ for a first-order transition \cite{Jorg2010}. In practice, quantum annealing including reverse annealing is performed diabatically and in a thermal environment, so that the results presented here may not carry over directly to practical situations.  

A further interesting question is to what extent quantum effects play an essential role in the present problem. A convenient classical model to be contrasted with the quantum model is spin-vector Monte Carlo (SVMC)~\cite{Shin2014}, in which we replace the Pauli matrices in the Hamiltonian, e.g. Eq.~(\ref{eq:hamiltonian1}), by classical unit vectors. As detailed in Appendix~\ref{appendix:C}, the resulting free energy has the same expression in the zero-temperature limit as its quantum counterpart, Eq.~(\ref{eq:free_energy_T=0}). Thus the phase  diagram remains the same. This feature has already been pointed out in the context of conventional QA~\cite{Susa2017}. Differences are expected to appear in dynamics, in particular when first-order transitions persist across the phase diagram as in Fig.~\ref{fig:PDNoRandomField}(a); classical dynamics at zero temperature are trapped in a local minimum since there is no classical mechanism for the system to go over the energy barrier, whereas quantum tunneling drives the system through the barrier. These aspects and other pertinent features of the quantum dynamics will be discussed in a forthcoming publication.

\begin{acknowledgments}
Part of the work was funded by the JSPS KAKENHI Grant No. 26287086. The research is based upon work partially supported by the Office of the Director of National Intelligence (ODNI), Intelligence Advanced Research Projects Activity (IARPA), via the U.S. Army Research Office contract W911NF-17-C-0050. The views and conclusions contained herein are those of the authors and should not be interpreted as necessarily representing the official policies or endorsements, either expressed or implied, of the ODNI, IARPA, or the U.S. Government. The U.S. Government is authorized to reproduce and distribute reprints for Governmental purposes notwithstanding any copyright annotation thereon.
\end{acknowledgments}
\appendix

\section{Static approximation}
\label{appendix:A}

Here, we calculate the partition function $Z=\mathrm{Tr}\exp(-\beta \hat{H})$ for the Hamiltonian Eq. (\ref{eq:hamiltonian1}) following Ref.~\cite{Seki2012}. We first use the Suzuki-Trotter formula, and the partition function can be written as 
\begin{align}
Z&=\lim_{M\to\infty}\mathrm{Tr}\left(e^{-(\beta/M)\big(s\hat{H}_0+(1-s)(1-\lambda)\hat{H}_{\rm init}\big)}e^{-(\beta/M)(1-s)\lambda\hat{V}_{\rm TF}}\right)^M \nonumber \\
&=\lim_{M\to\infty}\sum_{\{\sigma_i^z\}}\bra{\{\sigma_i^z\}}\Bigg\{\exp\left[\frac{\beta sN}{M}\left(\frac{1}{N}\sum_{i=1}^N\hat{\sigma}_i^z\right)^p+\frac{\beta s}{M}\sum_{i=1}^Nh_i\hat{\sigma}_i^z+\frac{\beta(1-s)(1-\lambda)}{M}\sum_{i=1}^N\epsilon_i\hat{\sigma}_i^z\right] \nonumber \\
&\times\exp\left[\frac{\beta(1-s)\lambda}{M}\sum_{i=1}^N\hat{\sigma}_i^x\right]\Bigg\}^M\ket{\{\sigma_i^z\}},
\end{align}
where $M$ is the Trotter number, and $\ket{\{\sigma_i^z\}}$ denotes an orthonormal basis that diagonalizes the $z$-component of the Pauli matrices. The summation is taken over all the possible basis states. We introduce $N$ closure relations $\hat{1}(\alpha)=\sum_{\{\sigma_i^z(\alpha)\}}\ket{\{\sigma_i^z(\alpha)\}}\bra{\{\sigma_i^z(\alpha)\}}$, where $\alpha=1,...,M$. Then, we have
\begin{align}
Z&=\sum_{\{\sigma_i^z(\alpha)\}}\prod_{\alpha=1}^M\exp\left[\frac{\beta sN}{M}\left(\frac{1}{N}\sum_{i=1}^N\sigma_i^z(\alpha)\right)^p+\frac{\beta s}{M}\sum_{i=1}^Nh_i\sigma_i^z(\alpha)+\frac{\beta(1-s)(1-\lambda)}{M}\sum_{i=1}^N\epsilon_i\sigma_i^z(\alpha)\right] \nonumber \\
&\times\prod_{\alpha=1}^M\bra{\{\sigma_i^z(\alpha)\}}\exp\left[\frac{\beta(1-s)\lambda}{M}\sum_{i=1}^N\sigma_i^x(\alpha)\right]\ket{\{\sigma_i^z(\alpha+1)\}}, %
\end{align}
where periodic boundary conditions are imposed, $\sigma_i^z(1)=\sigma_i^z(M+1)$ for $i=1,...,N$. Next, we introduce the following integral representation of the delta function:
\begin{align}
\delta\left(Nm-\sum_{i=1}^N\sigma_i^z\right)=\int d\tilde{m}\exp\left[-\tilde{m}\left(Nm-\sum_{i=1}^N\sigma_i^z\right)\right],
\end{align}
where $m$ denotes the magnetization (order parameter), and $\tilde{m}$ is the conjugate variable. We can rewrite $Z$ as
\begin{align}
\label{eq:Z}
Z&=\prod_{\alpha=1}^M\int\dots\int dm(\alpha)\,d\tilde{m}(\alpha)\exp\left[N\sum_{\alpha=1}^M\left(s\frac{\beta}{M}[m(\alpha)]^p-\tilde{m}(\alpha)m(\alpha)\right)\right]\nonumber \\
&\times\exp\left[\sum_{i=1}^N\ln\mathrm{Tr}\prod_{\alpha=1}^M\exp\left(\frac{\beta(1-s)\lambda}{M}\sigma_i^x(\alpha)+\tilde{m}\sigma_i^z(\alpha)+\frac{\beta s}{M}h_i\sigma_i^z(\alpha)+\frac{\beta(1-s)(1-\lambda)}{M}\epsilon_i\sigma_i^z(\alpha)\right)\right].
\end{align}
For $N\gg 1$, the saddle point condition can be imposed for $m(\alpha)$ as
\begin{align}
\tilde{m}(\alpha)=\frac{\beta}{M}ps[m(\alpha)]^{p-1}.
\end{align}
We use this equation and rearrange the exponent in Eq. (\ref{eq:Z}) as
\begin{align}
Z=\prod_{\alpha=1}^M\int\dots\int dm(\alpha)\,d\tilde{m}(\alpha)\exp\left[-N\beta f(\beta,s,\lambda;m(\alpha))\right],
\end{align}
where
\begin{align}
&f(\beta,s,\lambda;m(\alpha))=(p-1)sm^p \nonumber \\
&-\frac{1}{\beta}\sum_{i=1}^N\ln\mathrm{Tr}\exp\left(\beta\sum_{\alpha=1}^M\left(\left(\frac{ps}{M}[m(\alpha)]^{p-1}+\frac{sh_i}{M}+\frac{(1-s)(1-\lambda)\epsilon_i}{M}\right)\sigma_i^z(\alpha)+\frac{(1-s)\lambda}{M}\sigma_i^x(\alpha)\right)\right).
\end{align}
We now use the static approximation, which removes all the $\alpha$ dependence of the parameters, and finally obtain the free energy 
\begin{align}
f(\beta,s,\lambda;m)&=(p-1)sm^p-\frac{1}{\beta}\sum_{i=1}^N\ln\mathrm{Tr}\exp\beta\left(\left(psm^{p-1}+sh_i+(1-s)(1-\lambda)\epsilon_i\right)\sigma_i^z+(1-s)\lambda\sigma_i^x\right)\nonumber \\
&=(p-1)sm^p-\frac{1}{\beta}\left[\ln2\cosh\beta\sqrt{\left(psm^{p-1}+sh_i+(1-s)(1-\lambda)\epsilon_i\right)^2+(1-s)^2\lambda^2}\right]_i,
\end{align}
where $[...]_i$ is the configuration average with respect to the distribution of random field $h_i$.  This is Eq.~(\ref{eq:free_energy_Tfinite}) of the main text.

\section{Two transitions in the presence of bimodal random field}
\label{appendix:B}

\begin{figure*}[thb]
\centering
    \begin{tabular}{c}
      \begin{minipage}{0.5\linewidth}
			\centering
          \includegraphics[width=8cm,clip]{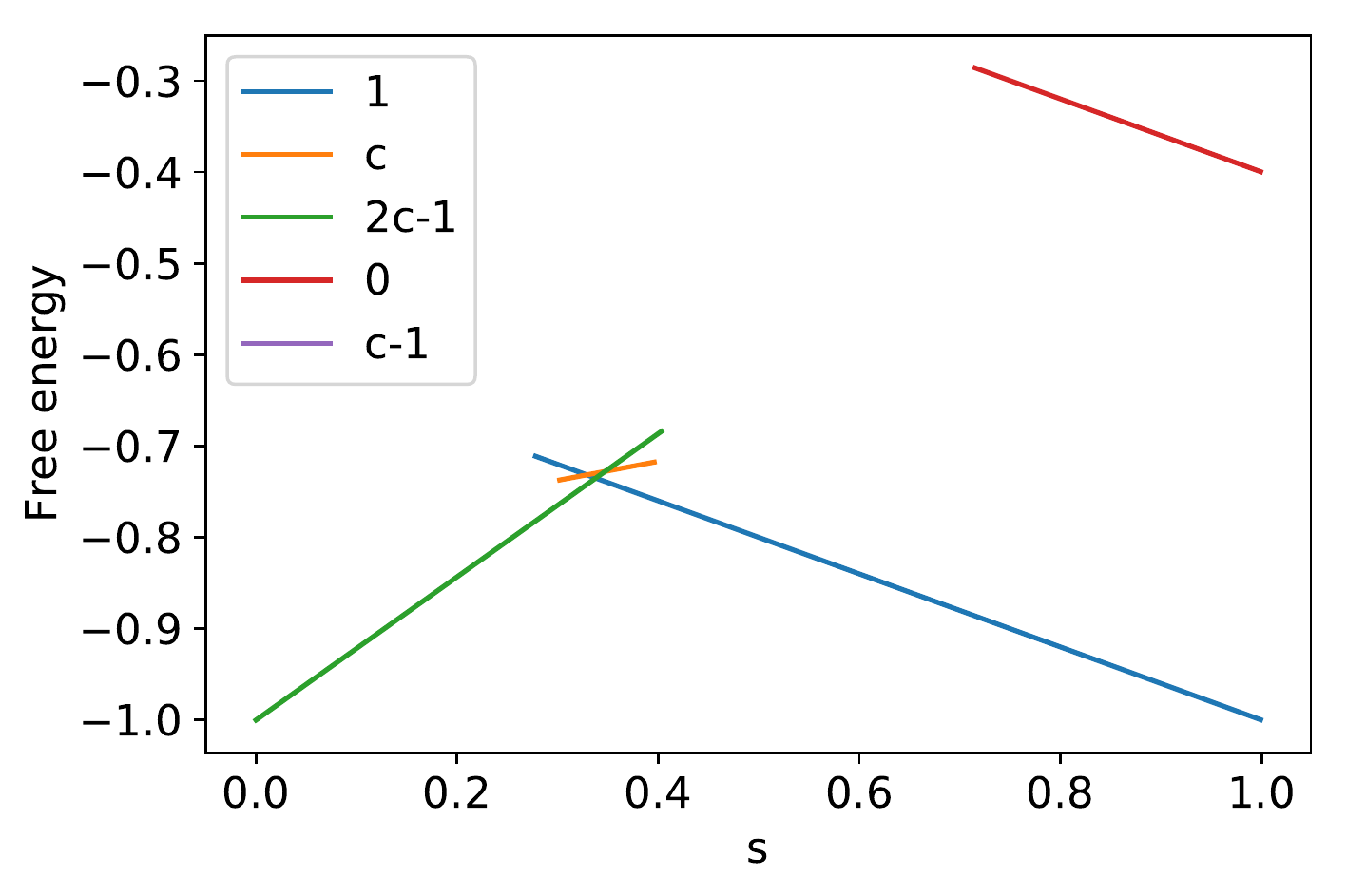}
          \hspace{1.6cm} (a)$\;h_0=0.4$
      \end{minipage}
      \begin{minipage}{0.5\linewidth}
        \begin{center}
          \includegraphics[width=8cm,clip]{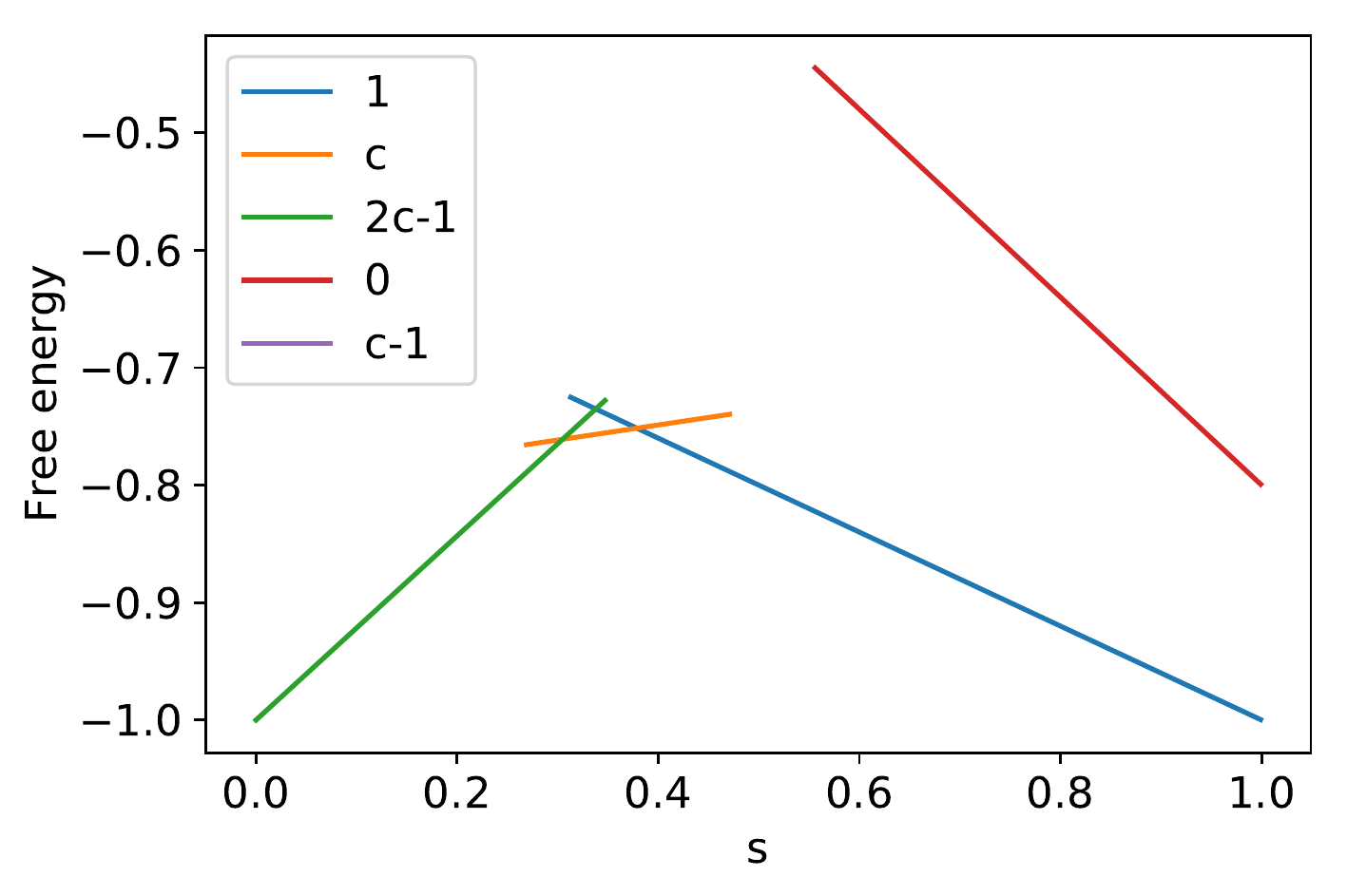}
          \hspace{1.6cm} (b)$\;h_0=0.8$
        \end{center}
      \end{minipage}
    \end{tabular}
    \caption{(Color online) Free energy of Eq.~(\ref{eq:free_energy_binary}) as a function of $s$ for each magnetization value $m$. The parameter $\lambda$ is fixed to 0, and $p=3$ and $c=0.8$. For these parameters, there is no region where the solution $m=c-1$ appears, and the corresponding curve does not appear in the figure. The magnetization having the smallest free energy is selected in the ground state.}
\label{fig:Lambda0FreeEnergy}
\end{figure*}

We show that, under a random bimodal distribution of local fields, there exist two first-order transitions at and near $\lambda=0$ as a function of $s$.  When $\lambda=0$, Eq.~(\ref{eq:self_consist_binary}) can be written as
\begin{align}
\label{eq:self_consist_binary2}
m=&\frac{c}{2}+\frac{1-c}{2}\mbox{sgn}\left(spm^{p-1}+sh_0-(1-s)\right) \nonumber \\
&+\frac{c}{2}\mbox{sgn}\left(spm^{p-1}-sh_0+(1-s)\right)+\frac{1-c}{2}\mbox{sgn}\left(spm^{p-1}-sh_0-(1-s)\right).
\end{align}
The solutions to this equation are $m=1$, $c$, $2c-1$, $0$, and $c-1$. The range of existence of each value is determined by the arguments of the sign functions in Eq.~(\ref{eq:self_consist_binary2}). The free energy for each magnetization value is shown in Fig. \ref{fig:Lambda0FreeEnergy}. The case of $h_0=0.4$ has one discontinuous transition from $m=2c-1$ to $m=1$ around $s=0.35$. In contrast,  if $h_0=0.8$, two discontinuous transitions exist from $m=2c-1$ to $m=c$ and from $m=c$ to $m=1$, around $s=0.3$ and $s=0.4$, respectively. In this way, transitions may occurs once or twice depending on the parameter $h_0$.
\section{Spin-vector Monte Carlo}
\label{appendix:C}

In this Appendix, we solve the equilibrium statistical mechanics of the spin-vector Monte Carlo (SVMC) model \cite{Shin2014} for the Hamiltonian of reverse annealing. In SVMC, we replace $\hat{\sigma}_i^z$ by $\cos\theta_i$ and $\hat{\sigma}_i^x$ by $\sin\theta_i$ in the Hamiltonian Eq.~(\ref{eq:hamiltonian1}) to have
\begin{align}
\hat{H}(s,\lambda)&=-s\left(N\left(\frac{1}{N}\sum_{i=1}^N\hat{\sigma}_i^z\right)^p+\sum_{i=1}^Nh_i\hat{\sigma}_i^z\right)-(1-s)\lambda\sum_{i=1}^N\hat{\sigma}_i^x-(1-s)(1-\lambda)\sum_{i=1}^N\epsilon_i\hat{\sigma}_i^z \nonumber \\
&\mapsto-s\left(N\left(\frac{1}{N}\sum_{i=1}^N\cos\theta_i\right)^p+\sum_{i=1}^Nh_i\cos\theta_i\right)-(1-s)\lambda\sum_{i=1}^N\sin\theta_i-(1-s)(1-\lambda)\sum_{i=1}^N\epsilon_i\cos\theta_i.
\end{align}
Let us evaluate the partition function $Z=\mathrm{Tr}\, \exp(-\beta H)$, where Tr means integrals over $0\le \theta_i<2\pi$. The magnetization $m = \sum_{i=1}^N \langle \hat{\sigma}_i^z\rangle /N$ is replaced by $m = (1/N)\sum_{i=1}^N \cos\theta_i$. If we introduce the integral representation of the delta function,
\begin{align}
\delta\left(Nm-\sum_{i=1}^N\cos\theta_i\right)=\int d\tilde{m}\exp\left[-\tilde{m}\left(Nm-\sum_{i=1}^N\cos\theta_i\right)\right],
\end{align}
we can rewrite $Z$ as
\begin{align}
\label{eq:appB_Z}
Z&=\mathrm{Tr}\int dm\delta\left(Nm-\sum_i\cos\theta_i\right)\nonumber\\
&\exp\left\{\beta(s(Nm^p+\sum_ih_i\cos\theta_i)+(1-s)\lambda\sum_i\sin\theta_i+(1-s)(1-\lambda)\sum_i\epsilon_i\cos\theta_i)\right\} \nonumber \\
&=\mathrm{Tr}\int dm\int d\tilde{m}\nonumber\\
&\exp\left\{-\tilde{m}(Nm-\sum_i\cos\theta_i)+\beta(s(Nm^p+\sum_ih_i\cos\theta_i)+(1-s)\lambda\sum_i\sin\theta_i+(1-s)(1-\lambda)\sum_i\epsilon_i\cos\theta_i)\right\}. 
\end{align}
The saddle point condition for $m$ leads to
\begin{align}
\tilde{m}=\beta spm^{p-1}.
\end{align}
We proceed to carry out the integral,
\begin{align}
&\mathrm{Tr}\exp\left[\beta\left(spm^{p-1}\sum_i\cos\theta_i+s\sum_ih_i\cos\theta_i+(1-s)\lambda\sum_i\sin\theta_i+(1-s)(1-\lambda)\sum_i\epsilon_i\cos\theta_i\right)\right]\nonumber \\
=&\prod_i\int^{2\pi}_0d\theta_i\exp\left[\beta\left(spm^{p-1}\cos\theta_i+sh_i\cos\theta_i+(1-s)\lambda\sin\theta_i+(1-s)(1-\lambda)\epsilon_i\cos\theta_i\right)\right]\nonumber \\
=&\prod_i2\pi I_0\left(\beta\sqrt{(spm^{p-1}+sh_i+(1-s)(1-\lambda)\epsilon_i)^2+(1-s)^2\lambda^2}\right),
\end{align}
where $I_n(x)$ is the modified Bessel function of the first kind. 
We then finally have
\begin{align}
Z=\int dm\exp\left[-N\beta s(p-1)m^p+\sum_i\log\left(2\pi I_0\left(\beta\sqrt{(spm^{p-1}+sh_i+(1-s)(1-\lambda)\epsilon_i)^2+(1-s)^2\lambda^2}\right)\right)\right],
\end{align}
from which we obtain the free energy per site
\begin{align}
f=&s(p-1)m^p-\frac{1}{\beta N}\sum_i\log\left(2\pi I_0\left(\beta\sqrt{(spm^{p-1}+sh_i+(1-s)(1-\lambda)\epsilon_i)^2+(1-s)^2\lambda^2}\right)\right)\nonumber \\
=&s(p-1)m^p-\frac{1}{\beta}\left[\log\left(2\pi I_0\left(\beta\sqrt{(spm^{p-1}+sh_i+(1-s)(1-\lambda)\epsilon_i)^2+(1-s)^2\lambda^2}\right)\right)\right]_i.
\end{align}
Let us consider low-temperature limit $\beta\to\infty$. Using the asymptotic form of the modified Bessel function,
\begin{align}
    I_0(\beta x)\approx\frac{e^{\beta x}}{\sqrt{2\pi\beta x}} ~(\beta x\gg 1).
\end{align}
Thus:
\begin{align}
f\to s(p-1)m^p-\left[z+\frac{1}{\beta}\log\left(\sqrt{2\pi}\left(z^{-1/2}+\frac{z^{-3/2}}{8}+\dots\right)\right)\right]_i,
\end{align}
where $z=\sqrt{(spm^{p-1}+sh_i+(1-s)(1-\lambda)\epsilon_i)^2+(1-s)^2\lambda^2}$. Taking the zero temperature limit $\beta\to\infty$, this coincides 
with Eq.~(\ref{eq:free_energy_T=0}) for the quantum model.
%

\end{document}